\documentclass[fleqn,usenatbib]{mnras}

\usepackage{newtxtext,newtxmath}

\usepackage[T1]{fontenc}
\usepackage{ae,aecompl}

\usepackage{graphicx}	
\usepackage{amsmath}	
\usepackage{rotating}
\usepackage{tabularx}

\usepackage{pdflscape} 
\usepackage{longtable}
\usepackage{newtxtext,newtxmath}

\title[\textit Quenching bursting and galaxy morphologies]{Quenching, bursting and galaxy shapes: colour transformation as a function of morphology}

\author[C. de S\'a-Freitas et al.]{
C. de S\'a-Freitas,$^{1,2}$\thanks{E-mail: camila.desafreitas@eso.org}
T. S. Gon\c{c}alves,$^{1}$ R.R. de Carvalho,$^{3}$ K. Menéndez-Delmestre,$^{1}$ \newauthor P.H. Barchi,$^{4}$ V. M. Sampaio,$^{3}$  Antara Basu-Zych,$^{5,6,7}$ Behnam Darvish,$^{8}$ Christopher Martin$^{8}$ \\
$^{1}$Observat\'orio do Valongo, Universidade Federal do Rio de Janeiro, Ladeira do Pedro Antônio, 43 - Centro, Rio de Janeiro - RJ, 20080-090, Brazil\\
$^{2}$European Southern Observatory, Karl-Schwarzschild-Straße 2, 85748 Garching bei München, Germany\\
$^{3}$NAT-Universidade Cruzeiro do Sul / Universidade Cidade de São Paulo, Rua Galvão Bueno, 868, 01506-000, São Paulo, SP, Brazil \\
$^{4}$Instituto Nacional de Pesquisas Espaciais/MCT, São José dos Campos, Brazil \\
$^{5}$ NASA Goddard Space Flight Center, Code 662, Greenbelt, MD 20771, USA \\
$^{6}$Center for Space Science and Technology, University of Maryland Baltimore County, 1000 Hilltop Circle, Baltimore, MD 21250, USA \\
$^{7}$Department of Physics, University of Maryland Baltimore County, Baltimore, MD 21250, USA \\
$^{8}$Cahill Center for Astrophysics, California Institute of Technology, 1216 East California Boulevard, Pasadena, CA 91125, USA
}

\date{Accepted XXX. Received YYY; in original form ZZZ}

\pubyear{2020}

\begin{document}
\label{firstpage}
\pagerange{\pageref{firstpage}--\pageref{lastpage}}
\maketitle

\begin{abstract}

Different mechanisms for quenching star formation in galaxies are commonly invoked in the literature, but the relative impact of each one at different cosmic epochs is still unknown. In particular, the relation between these processes and morphological transformation remains poorly understood. In this work, we measure the effectiveness of changes in star formation rates by analysing a new parameter, the Star Formation Acceleration (SFA), as a function of galaxy morphology. This methodology is capable of identifying both \textit{bursting} and \textit{quenching} episodes that occurred in the preceding 300 Myrs. We use morphological classification catalogs based on Deep learning techniques. Our final sample has $\sim$14,200 spirals and $\sim$2,500 ellipticals. We find that elliptical galaxies in the transition region have median shorter quenching timescales ($\tau< 1$ Gyr) than spirals ($\tau\geq 1$ Gyr). This result conforms to the scenario in which major mergers and other violent processes play a fundamental role in galaxy evolution for most ellipticals, not only quenching star formation more rapidly but also playing a role in morphological transformation. We also find that $\sim$two thirds of galaxies bursting in the green valley in our sample are massive spirals ($M_\star \geq 10^{11.0}M_\odot$) with signs of disturbance. This is in accordance with the scenario where low mass galaxies are losing their gas in a interaction with a massive galaxy: while the former is quenching, the last is being refueled and going through a burst, showing signs of recent interaction. 

\end{abstract}




\section{Introduction}
\label{Chapter1}


Galaxies have long been classified according to their apparent shape, ever since the seminal work by \cite{hubble1927classification} in the optical. Many authors have later systematically classified galaxy optical morphologies according to an array of structural and visual parameters (\citealp{sandage1961hubble} - \citeyear{sandage1995book}; \citealp{de1959classification}; \citealp{morgan1958preliminary} - \citeyear{morgan1970spiral}; \citealp{van1960preliminary} - \citeyear{van1976new}). Although these works do not take into account any additional information other than shape, many studies have found that there are correlations between morphology and other physical properties (e.g.
\citealp{roberts1994physical}; \citealp{blanton2009physical}, \citealp{nair2010catalog}). Star formation rates (SFR), in particular, seem to correlate strongly with galaxy morphologies, with spiral galaxies hosting younger stellar populations and actively forming new stars, while elliptical galaxies are on the most part passive, with older stellar populations and few recently formed stars (e.g. \citealp{kauffmann2003stellar}).

This dichotomous behavior can be easily visualized in a color-magnitude diagram (CMD) with its well-known bimodality, where the \textit{blue cloud} is mostly populated by star-forming spirals and the \textit{red sequence} is mostly populated by passive ellipticals (e.g. \citealp{baldry2004quantifying}; \citealp{schawinski2014green}; \citealp{bremer2018galaxy}). In between these two regions one can define the \textit{green valley} (e.g. \citealp{salim2007uv}; \citealp{martin2007uv}; \citealp{wyder2007uv}), thus called for the relative dearth of objects between two population peaks and commonly regarded as a transition region between star-forming and passive galaxies (e.g. \citealp{martin2007uv}; \citealp{gonccalves2012quenching}; \citealp{nogueira2018star}), with on average older stellar populations than the blue cloud and younger than the red sequence, as demonstrated by \cite{pan2013green}. Consequently, processes responsible for the transition of galaxies through the \textit{green valley} should somehow correlate with a morphological transformation for most galaxies, since the stellar populations are younger (spirals) versus older (ellipticals).

Many processes are invoked by authors as responsible for quenching star formation and modifying galaxy colors from bluer to redder colors. AGN and stellar feedback are often cited as possible mechanisms for expelling the gas responsible for star formation in young clusters (e.g. \citealp{fabian2012observational}; \citealp{hopkins2014galaxies}) while secular processes could gradually exhaust the gas reservoir within the galaxy (e.g. \citealp{sheth2005secular};  \citealp{fang2013link};  \citealp{schawinski2014green}; \citealp{bluck2014bulge}; \citealp{nogueira2018star}). In dense environments, galaxies can also experience ram pressure, strangulation, galaxy interactions, galactic harassment, mergers, disk instabilities (e.g. \citealp{boselli2014origin}) or halo quenching (e.g. \citealp{birnboim2003virial}), all of which impact the gas reservoir of the galaxy or the long-term availability of gas in the galactic vicinity. This, in turn, has a strong effect on star formation histories. Furthermore, different processes can be associated with different quenching timescales, which can be distinguished between fast ($<$ 300 Myrs) and slow ($\gtrapprox$ 1 Gyr) processes (\citealp{schawinski2007effect}; \citealp{gonccalves2012quenching}; \citealp{martin2017quenching}).

Mergers in particular have long been quoted as one of the main drivers of rapid transition across the CMD and morphological change (e.g. \citealp{schawinski2014green}, \citealp{nogueira2018star}). Major mergers, where the masses of the two merging galaxies are comparable, can lead to hydrodynamical instabilities that affect the galaxy in many ways: simulations have shown that mergers can trigger starbursts and lead to gas exhaustion, followed by rapid quenching (e.g. \citealp{di2005energy}). For example, these interactions can expel the gas through conservation of angular momentum (\citealp{van1990nuclear}). 

Likewise, mergers can also drive gas inflows to the center of the galaxy, feeding a phase of rapid AGN growth (e.g. \citealp{cattaneo2005spectral}; \citealp{king2005hierarchical}). In its turn, AGN feedback is able to quench star formation, either mechanically (expelling gas) or energetically (heating the ISM), causing a drop in star formation rates and ensuring no subsequent star formation episodes and disk formation from gas reaccretion (e.g. \citealp{croton2006many}; \citealp{schawinski2009moderate}; \citealp{fabian2012observational}; \citealp{dubois2016}; \citealp{nogueira2019compact}). The investigation of Ultraluminous Infrared Galaxies (ULIRGs), in particular, emphasizes the role of strong interactions and major/minor mergers as agents responsible for triggering the quasar stage in galaxies' central SMBH (eg \citealp{sanders1990ultraluminous}; \citealp {schweitzer2006spitzer}; \citealp {veilleux2006deep}). 

The opposite can also happen, i.e, mergers may not only quench but also trigger star formation (e.g. \citealp{rampazzo2007galaxy}; \citealp{thilker2010ngc}; \citealp{thomas2010environment}; \citealp{salim2012galaxy}; \citealp{fang2013link}). Indeed, many authors have shown that a significant fraction of \textit{green valley} galaxies are not currently quenching, but undergoing a process of rejuvenation (e.g. \citealp{rampazzo2007galaxy}; \citealp{martin2017quenching}; \citealp{pandya2017nature}), through accretion of a gas-rich companion. 

In addition, mergers can also affect galactic morphologies, leading to a bulge growth and transforming the entire structure of the galaxy (e.g. \citealp{toomre1972galactic}; \citealp{springel2005simulations}). Minor mergers, in turn, commonly defined as mergers between galaxies with mass ratio smaller than 1:3 (e.g. \citealp{miralles2014ionizing}) are expected to be more frequent (e.g. \citealp{lotz2011major}). These encounters can increase SFR (e.g. \citealp{kennicutt1987effects}; \citealp{martin2017quenching}) and trigger nuclear activity (e.g. \citealp{hopkins2006relation}). However, unlike major mergers, a single minor merger is much less dynamically violent, and does not affect the galaxy morphology as a whole, allowing the host to keep its global form, albeit with perturbations and asymmetries (e.g. \citealp{ruiz2020recurrent}). On the other hand, multiple minor mergers can also lead to significant morphological transformation (e.g. \citealp{bournaud2007multiple}) and, eventually, to morphological quenching (e.g. \citealp{martig2009morphological}; \citealp{martin2018role})

In an attempt to quantify the effectiveness of quenching mechanisms, \cite{martin2017quenching} introduced a new non-parametric methodology to measure the instantaneous colour variation in galaxies: the Star Formation Acceleration (SFA), which can be understood as the instantaneous time derivative of the star formation rate (SFR) averaged over the past 300 Myr. The authors found that, although the net mass flow is towards the red sequence (in accordance with \citealp{rampazzo2007galaxy}; \citealp{thilker2010ngc}; \citealp{thomas2010environment}; \citealp{salim2012galaxy}), a significant fraction of green valley galaxies are bursting, i.e. these galaxies are actually becoming bluer, with increasing SFR. Splitting the sample by mass, the authors attributed most of the bursting activity to massive galaxies, in a scenario in which molecular gas from satellites is falling onto the central galaxy, quenching star formation in the former and triggering in the latter. Using the same methodology, \cite{darvish2018quenching} showed that most of the quenching (bursting) of star formation happens in low (high) stellar mass galaxies, in particular those located in dense (less dense) environments, highlighting the combined effects of mass and environment in the transformation of galaxies since $z~\sim~1$.

All of the aforementioned results combined show that galaxy evolution is extremely complex, and the result of a combination of a number of different physical processes. Furthermore, the colour-morphology relation also strongly indicates that transformations in galaxy shape must be strongly intertwined with changes in star formation rates. Nevertheless, previous works attempting to correlate galaxy morphology in the green valley with star formation timescales rely on more simplistic star formation history models that only account for quenching, and no bursting (e.g. \citealp{schawinski2014green}; \citealp{nogueira2018star}). Therefore, the SFA methodology introduced by \cite{martin2017quenching} presents an excellent opportunity to investigate this process without any preliminary assumptions on whether a galaxy in the green valley is increasing or decreasing its star formation rates.

In order to investigate evolutionary pathways for different morphologies and understand whether or not physical processes can affect galaxy structure, in this work we apply the methodology presented by \cite{martin2017quenching} and analyse typical SFA behaviour for spiral vs. elliptical galaxies. Our main goal is to determine whether galaxies of different morphologies are evolving at different rates, in an attempt to correlate quenching and bursting timescales with morphological transformations that need to occur to reproduce the shapes of these galaxy populations.


In Section \ref{Chapter2} we describe our samples and the different morphological classifications and we present the methodology used to determine SFA; in Section \ref{Chapter4} we show our results, discussing their implications in Section \ref{Chapter5}; finally, we summarize our findings in Section \ref{Summary}. This work uses standard cosmology throughout the paper, with $\Omega_{M}=0.3$, $\Omega_{\Lambda}=0.7$ and $h = 0.7$.

\section{Sample \& Methodology}
\label{Chapter2}

\subsection{Galaxy Sample}
\label{sub_sec_galaxysample}

Our original sample of $\sim$750,000 galaxies was selected from the Sloan Digital Sky Survey (SDSS) Data Release 12 (\citealp{alam2015eleventh}). We also limited our sample to galaxies brighter than Petrosian r-band magnitude $m_{p,r} = 17.77$, as a threshold for spectroscopic completeness (\citealp{strauss2002spectroscopic}). In order to determine to which absolute magnitude this corresponds, we follow the procedure described in \cite{la2010spider}, which accounts for variations caused, for instance, by k-corrections and intrinsic reddening, restricting our selection to galaxies between redshifts $0.05 \leq z \leq 0.095$. To that end, we consider $95\%$ completeness in each observed magnitude bin, and use a linear fit to determine its intersection with the observational limit of the spectroscopic sample, which is measured to be M$_{lim,90\%} = -20.42$, i.e, we infer that our sample is at least $95\%$ complete out to the limiting redshift of $z=0.095$. We matched this complete sample of $\sim 72,000$ objects with the \textit{Galaxy Explorer Survey - General Release 6} (\citealp{martin2005galaxy}) -- GALEX GR6, within a 3'' maximum distance. Finally, we adopt SFA values from \cite{martin2017quenching}. We note that their sample requires detection in the GALEX \textit{Medium Imaging Survey} -- MIS. Our final sample of galaxies comprises a total of $\approx 20,900$ objects with UV detections and SFA measurements.

\begin{figure}
        \centering
        \includegraphics[width=\columnwidth]{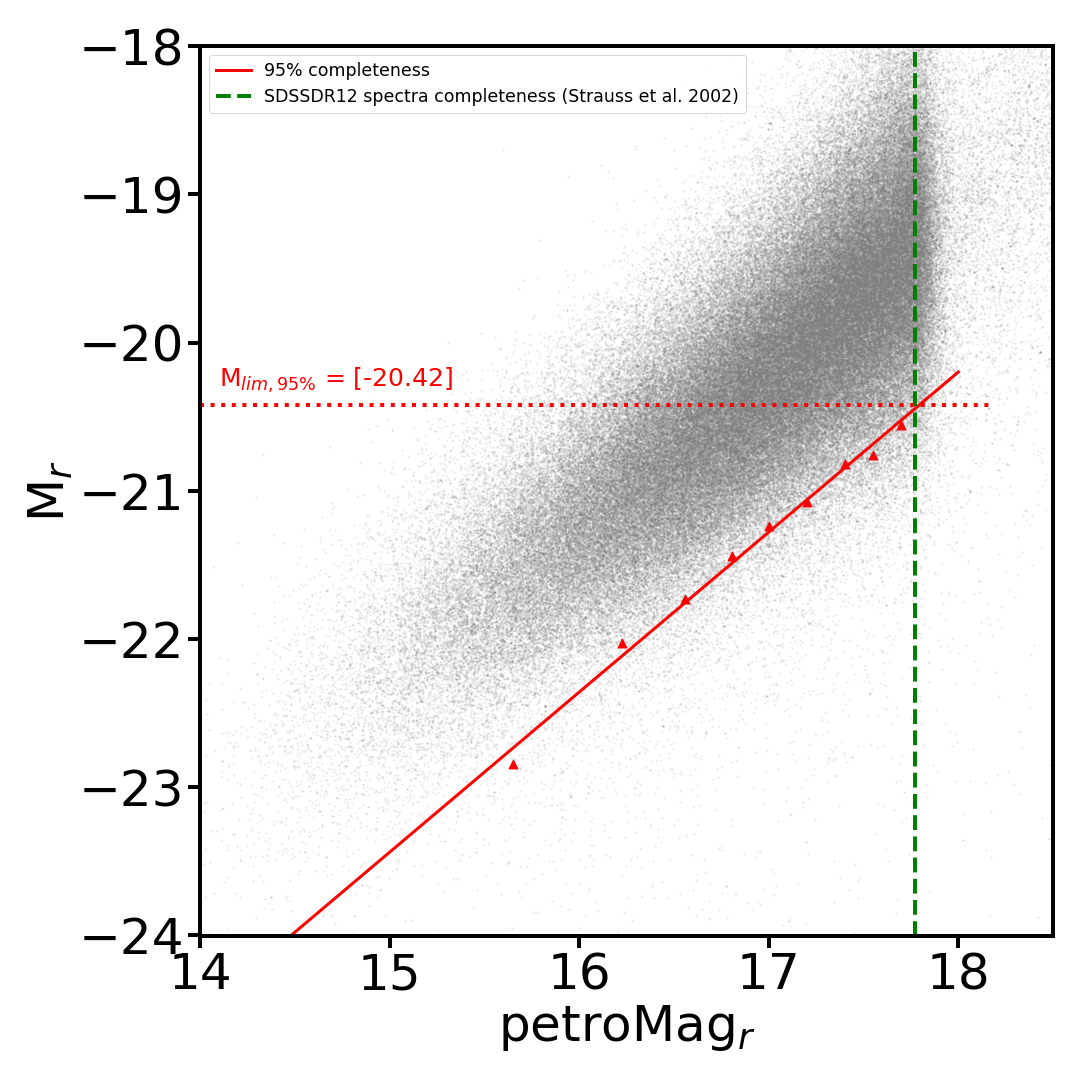}
        \caption{Apparent magnitude in the r-band \textit{vs.} absolute magnitude for the SDSS sample, for redshifts between $0.050 \leq z \leq 0.095$. The vertical dashed green line represents the Petrosian magnitude limit of $m_{p,r} = 17.77$ (\citealp{strauss2002spectroscopic}). The red line is a linear fit to a $95\%$ completeness limit in 9 apparent magnitude bins with same number of galaxies. Each bin is represented by a red circle. The horizontal red dotted line represents the point where the red line intercepts the vertical line, defining the absolute magnitude limit for which the sample is $95\%$ complete, $M_{lim} = -20.42$}
        \label{fig_malmquist}
\end{figure}




According to those criteria, our sample is complete for (\textit{i}) massive objects, with masses greater than $M_* \gtrsim 10^{10} M_\odot$, due to our absolute magnitude limit (Section \ref{sub_results_sfa_mass}); and (\textit{ii}) for colours bluer than $(NUV-r)_0 \sim 5$ due to our need for reliable UV measurements for the determination of SFA (Section \ref{sub_results_cmd} - Figure \ref{fig_CMD}).


\subsection{Dust attenuation}

Our main interest lies in the galaxies within the green valley. We correct for internal reddening, considering that it has been demonstrated that galaxies with high SFRs and high dust content may contaminate the green valley artificially, not representing an actual state of transitioning between populations (e.g. \citealp{sodre2013nature}; \citealp{gonccalves2012quenching}). To do so, we followed the \cite{calzetti2000dust} dust correction procedure for UV-optical colours, based on the Balmer decrement. 





High-resolution spectroscopy is necessary to correctly measure the Balmer emission lines, due to the presence of strong metal and Balmer absorption features. The measurement of hydrogen line flux ratios for galaxies with low star-formation rates (in particular faint H$\beta$ lines) is challenging, making the determination of extinction for all galaxies in our sample difficult by this direct method. We hence followed the methodology introduced by \citet{groves2011balmer} based on equivalent width (EW) measures and the theoretical unreddened line ratios  $H_\alpha$/$H_\beta$ \citep{calzetti1994dust,osterbrock2006astrophysics}. Briefly stated, \citet{groves2011balmer} used a large SDSS DR7 sample to empirically define the EW Balmer decrement as follows:

\begin{equation}
    \textrm{log({H}}_\alpha / \textrm{H}_\beta \textrm{) = log}\left(\frac{\textrm{EW(H}\alpha) + a_1}{{\textrm{EW(H}}\beta) + a_2}\right) + [-0.26 + 0.39(g-r)]
    \label{eq_ew1}
\end{equation}

\noindent $a_1$, $a_2$ $= 4.1$ and $4.4$ for EW(H$\alpha$)$\,\gtrsim 30$ \AA, and $3.5$, $3.8$ for EW(H$\alpha$)$\;<30$ \AA. With log(H$_\alpha$/H$_\beta$) we can calculate the colour excess as following (\citealp{calzetti1994dust}):


\begin{equation}
    E(B-V)_{gas} = 0.935 \log{\left(\frac{H_\alpha/H_\beta}{2.88}\right)},
    \label{eq:colour_excess_gas}
\end{equation}
where and $E(B-V)_{star} = 0.44E(B-V)_{gas}$. Lastly, with the colour excess and traditional emission lines, we can calculate the intrinsic magnitudes, corrected for internal extinction. Following  \cite{calzetti2000dust}: 

\begin{equation}
    m_{int}(\lambda) = m_{obs}(\lambda)~-k^e(\lambda)~E(B-V)_{star}
    \label{eq_nuvr0}
\end{equation}

\noindent where $k^e(\lambda)$ can be derived in equation (\ref{eq_ke}):

\begin{equation}
    k^e (\lambda) = 2.659~(-2.156+1.509/\lambda -0.198/\lambda^2 +0.011/\lambda^3) +R_v
    \label{eq_ke}
\end{equation}

\noindent for $\lambda_{NUV} = 0.2267 \mu m$\footnote{https://asd.gsfc.nasa.gov/archive/galex/Documents/ERO\_data\_description\_2.htm}, $\lambda_r = 0.6231 \mu m$\footnote{https://skyserver.sdss.org/dr1/en/proj/advanced/color/sdssfilters.asp} and $R_v = 4.05$ (\citealp{calzetti2000dust}).


We would like to emphasize that this procedure was only applied to galaxies with H$\alpha$ and H$\beta$ lines in emission (EW<0), located mostly in the blue cloud and \textit{apparent} green valley. We have assumed galaxies with no discernible equivalent width for emission lines show little to no reddening. Finally, for the K-correction we used the open code calculator based on redshift and color from \cite{chilingarian2010analytical} and \cite{chilingarian2011universal}\footnote{http://kcor.sai.msu.ru/getthecode/\#python}. Every $(NUV-r)_0$ color hereafter will be dust- and K-corrected to z = 0.

All magnitudes are also corrected for Milky Way extinction, with values provided by SDSS for optical colors (\textit{extinction\_u, extinction\_g, extinction\_r, extinction\_i, extinction\_z}), and by GALEX for NUV (\textit{e\_bv}).


\subsection{Morphological Classification}
\label{sub_morphological}

In order to investigate the evolutionary pathways for different morphologies, we used a morphological classification catalogue based on Deep Learning (\citealp{barchi2019machine}), with morphological information for 670,560 galaxies. The catalogue has values for non-parametric morphological parameters and morphological classification provided by methods from traditional machine learning and deep learning as well. Here, in this work, we use the classification obtained with the deep learning approach (see \citealp{barchi2019machine} for more details on the automated classification methods), considering 80$\%$ of confidence in the classification.

Since we are using the 80$\%$ criterion, we expect little contamination in the morphological classification between spirals and ellipticals; this provides us with a high level of purity in our sample.

This resulted in a final sample of $14,219$ spirals and $2,590$ ellipticals, each with its individual SFA value.



\subsection{\textit{Star Formation Acceleration} -- SFA}
\label{subsec_sfa}

In order to quantify whether a galaxy is \textit{quenching} or \textit{bursting}, we use the SFA parameter, introduced by \citealp{martin2017quenching}. SFA is a proxy for the instantaneous SFR time derivative. It is defined as:

\begin{equation}
    SFA \equiv \frac{d(NUV-i)_0}{dt},
    \label{eq_SFA}
\end{equation}
where $(NUV-i)_0$ is corrected for extinction, $dt$ corresponds to a time interval of the past 300 Myr and where SFA is measured in units of [Gyr$^{-1}$]. By definition, a positive SFA indicates that the galaxy has been \textit{quenching} for the past 300 Myr, whereas a negative value indicates that the galaxy has been \textit{bursting}. 

\cite{martin2017quenching} use semi-analytical models (\citealp{de2006formation}) coupled with the cosmological N-body \textit{Millenium} simulation (\citealp{springel2005simulations}) to generate a catalogue of simulated galaxies that lie within the redshift range $0~<~z~<~6$. Based on this catalogue of mock galaxies, physical properties such as SFR, \textit{Star Formation History} (SFH), stellar mass and SFA are known {\it a priori}.



Observables, including broadband colours and spectroscopic indices (e.g. $FUV-NUV$ $NUV-i$, $NUV-u$, $u-g$, $g-r$, $H_{\delta,A}$, $D_n(4000)$, $M_i$), on the other hand, are produced as output by using stellar population synthesis models. \citet{martin2017quenching} used models from \cite{bruzual2003stellar} and a mixed extinction model -- one that takes into account the starburst extinction law from \cite{calzetti2000dust} and the Milky Way extinction law (\citealp{cardelli1989relationship}) -- in order to produce synthetic spectra for each galaxy in the mock catalogue.

The simulated sample is then subdivided in bins of redshift and the strength of the 4000-\AA~ break in the galaxy spectra — based on its quantification by the $D_n(4000)$ index, as defined in \cite{kauffmann2003stellar}. Within each redshift bin and $D_n(4000)$ bin the authors perform linear regression fits between physical properties (e.g., SFR, SFH, SFA) and observables (e.g., $FUV-NUV$ $NUV-i$, $NUV-u$, $u-g$, $g-r$, $H_{\delta,A}$, $M_i$), generating multiple matrices of regression coefficients. This allows for later determination of physical properties from photometric and spectroscopic measurements of galaxies in observational surveys. The observables used were the rest-frame $FUV-NUV$, $NUV-u$, $u-g$, $g-r$, and $r-i$ colors, $M_i$ absolute magnitude, $D_n(4000)$, and $H_{\delta,A}$ (\citealp{martin2017quenching}; \citealp{darvish2018quenching}).

We used this methodology to infer SFA for the galaxy sample defined in Section \ref{Chapter2}. Although the determination of SFA is uncertain for individual galaxies, given the standard deviation of $\sim 1$ Gyr$^{-1}$ in the linear regression in \citet{martin2017quenching}, the behaviour for a statistically significant sample is robust. With this in mind, we will show our results as behaviour for individual colour bins of $\Delta(NUV-r) = 0.5$. For more details about the method, we refer the reader to \cite{martin2017quenching}.

We also note that, since any AGN contribution to the final SFA value is minimal (\citealp{martin2017quenching}), we do not exclude those objects from our sample.





\section{Results}
\label{Chapter4}
\label{chp4}

Our main objective in this work is to determine whether \textit{quenching} and \textit{bursting }timescales, as measured through the SFA index, show any correlation with galaxy morphologies. In this section, we present results from our analysis, also taking into consideration other properties such as galaxy colour and stellar mass.

\subsection{Color-magnitude diagram}
\label{sub_results_cmd}

\begin{figure*}
    \centering
    \includegraphics[width=2\columnwidth]{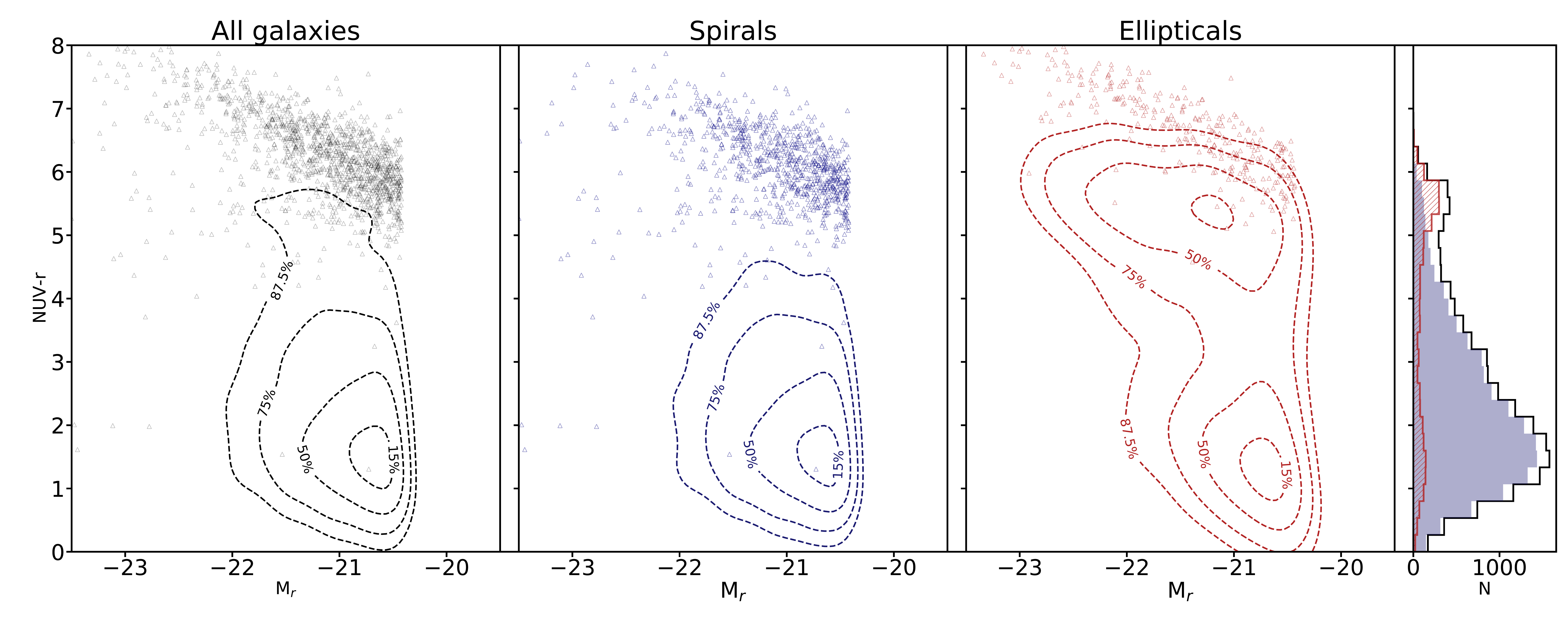}

    \caption{From left to right, color-magnitude diagrams for all galaxies with morphological classification (dashed black curves), spirals (blue dashed curves) and ellipticals (red dashed curves). Contours indicate sample percentiles within each level: 15$\%$, 50$\%$, 75$\%$ and 87.5$\%$. In the far-right, we show the $(NUV-r)_0$ distribution for all three samples, following the colour-code. We also show the $(NUV-r)_{0,lim}$, that is the ${NUV-r}_0$ lower limit, for faint-UV galaxies that are in the GALEX footprint but do not have UV-detections (triangles).}
    \label{fig_CMD}
\end{figure*}

\begin{figure*}
    \centering
    \includegraphics[width=0.65\columnwidth]{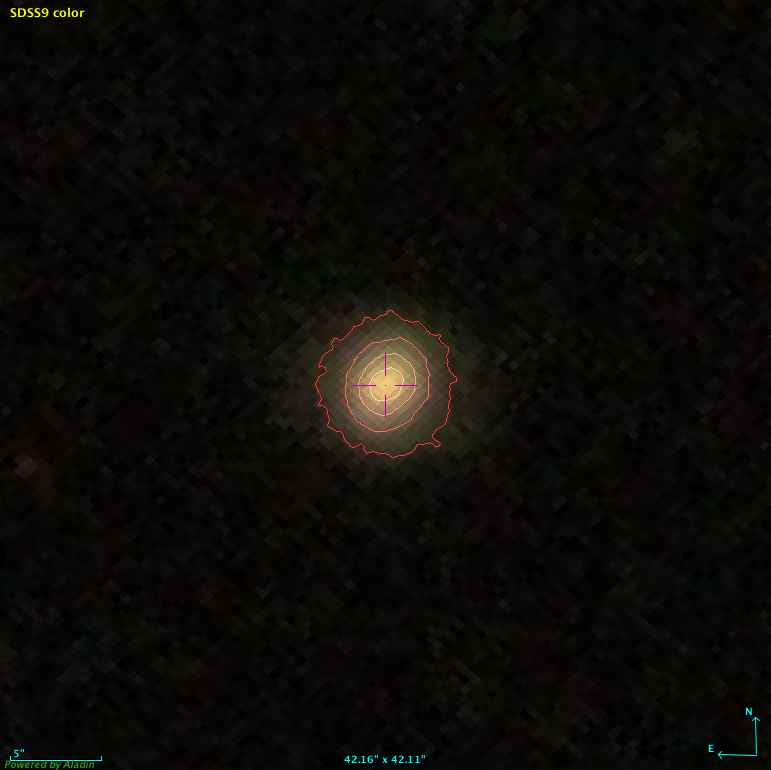}
    \includegraphics[width=0.65\columnwidth]{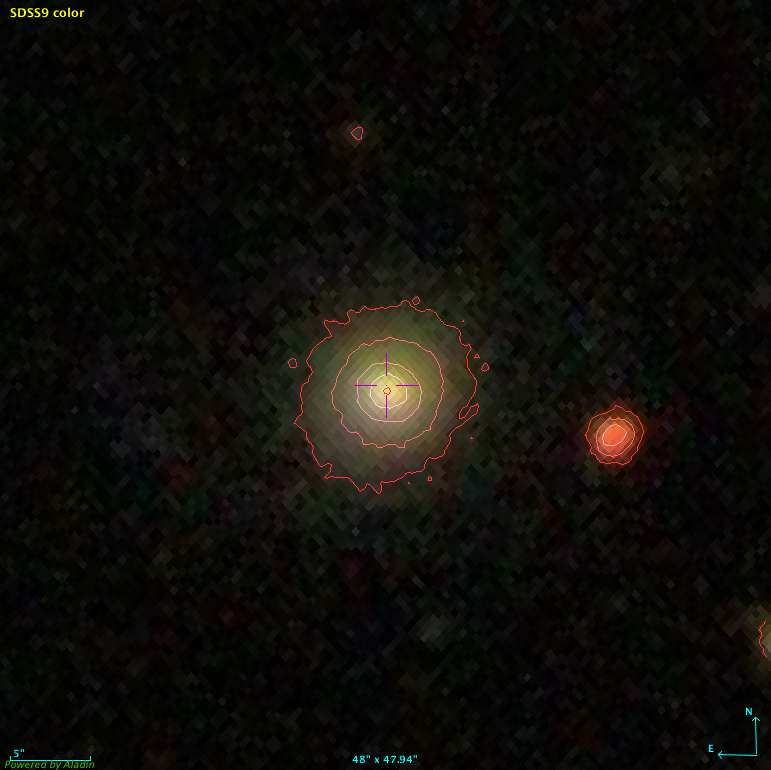}
    \includegraphics[width=0.65\columnwidth]{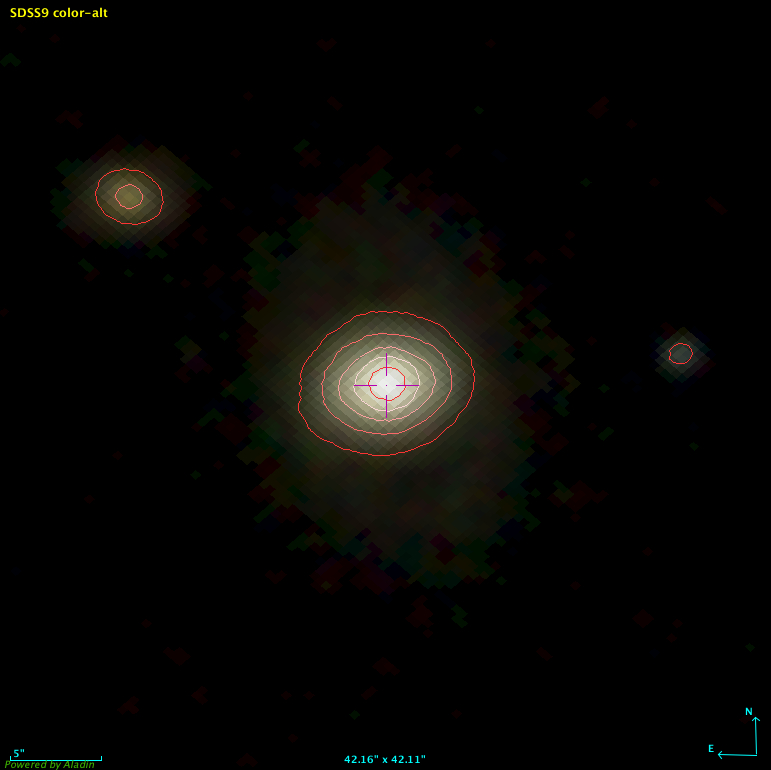}
    \includegraphics[width=0.65\columnwidth]{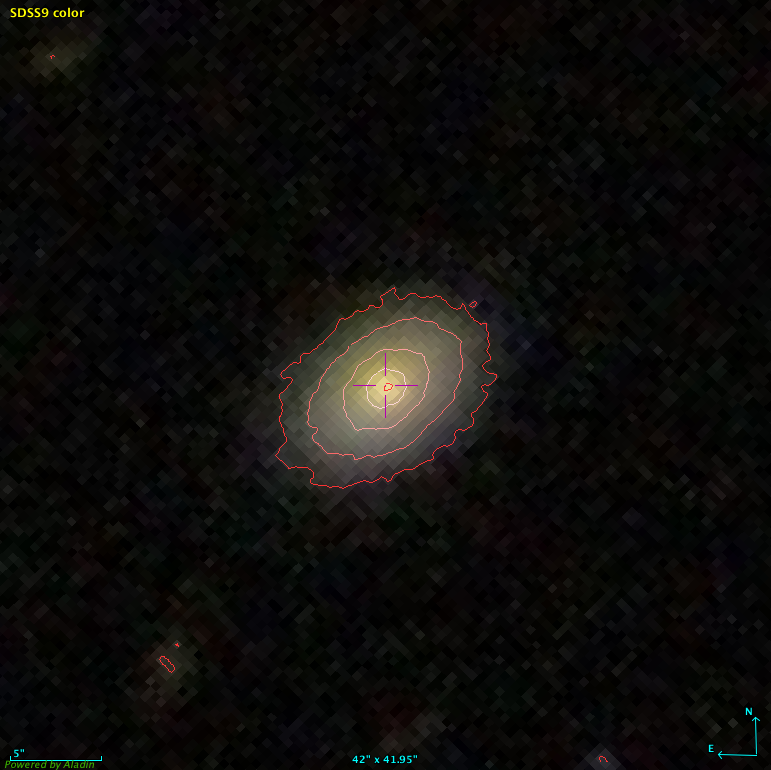}
    \includegraphics[width=0.65\columnwidth]{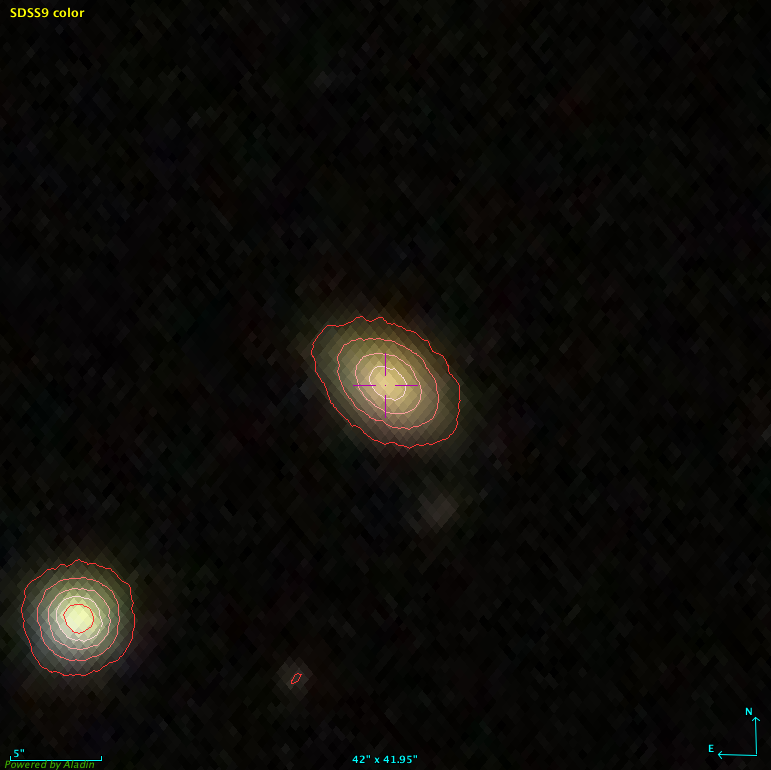}
    \includegraphics[width=0.65\columnwidth]{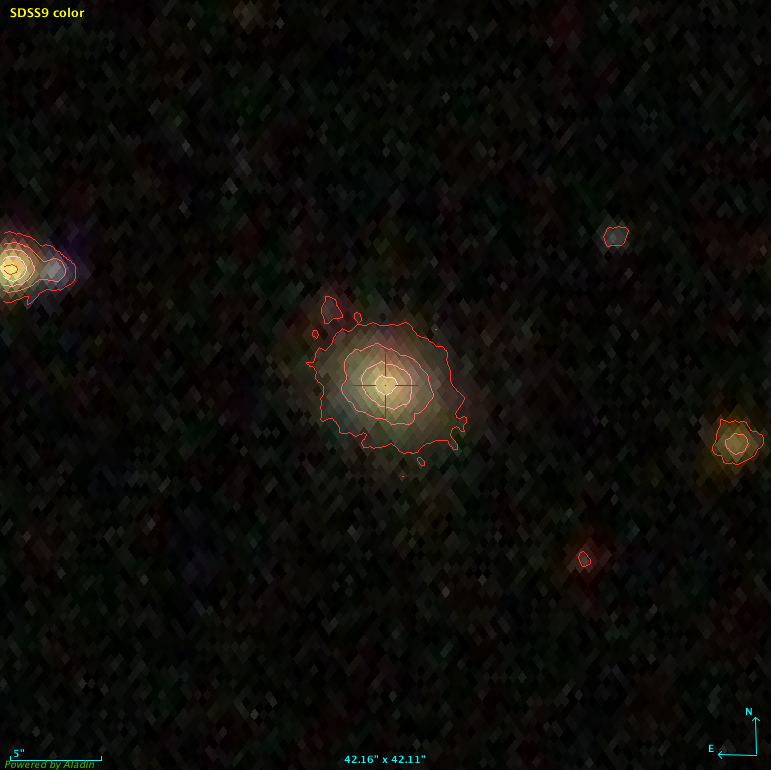}
    \includegraphics[width=0.65\columnwidth]{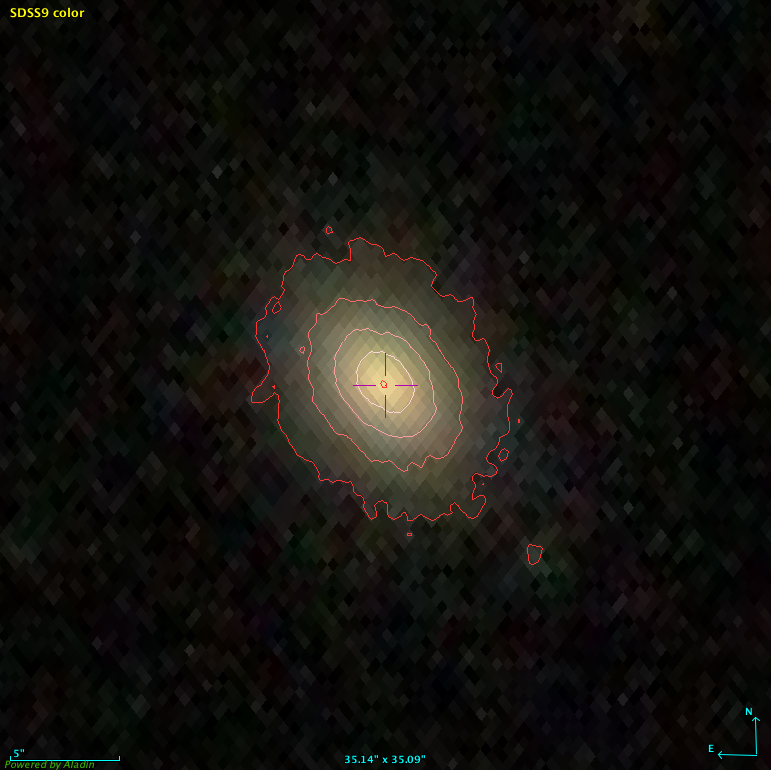}
    \includegraphics[width=0.65\columnwidth]{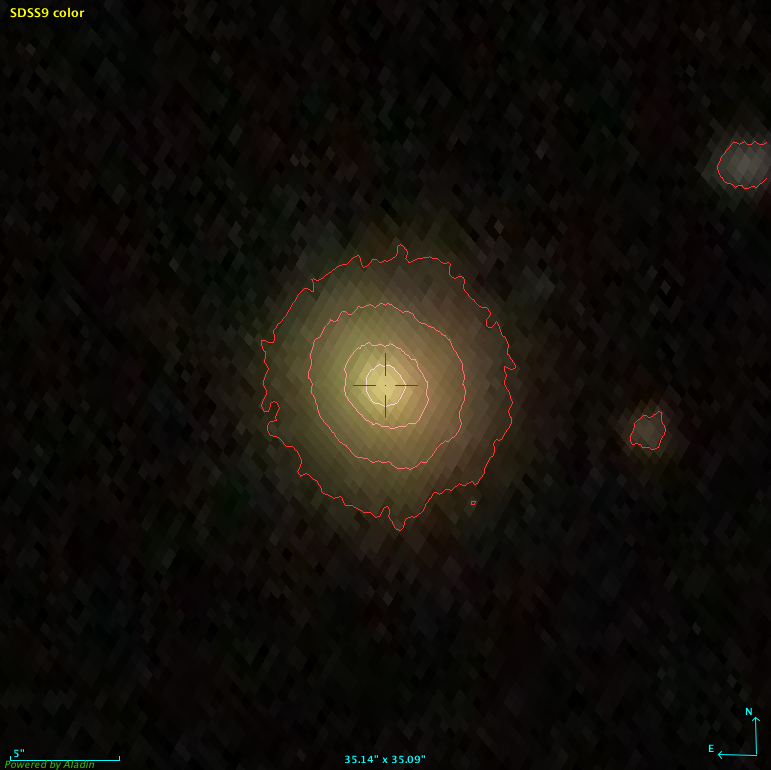}
    \includegraphics[width=0.65\columnwidth]{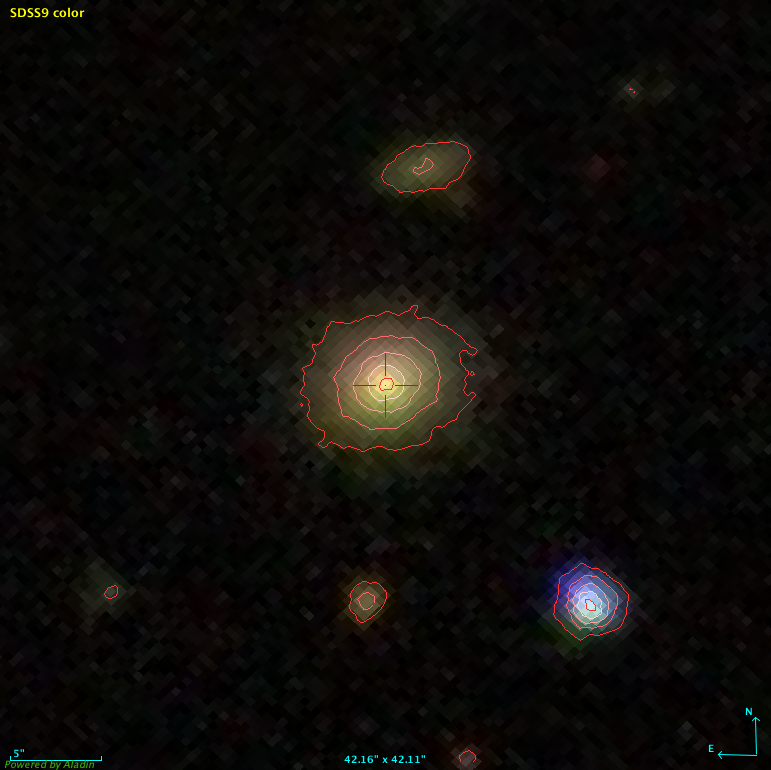}
    \includegraphics[width=0.65\columnwidth]{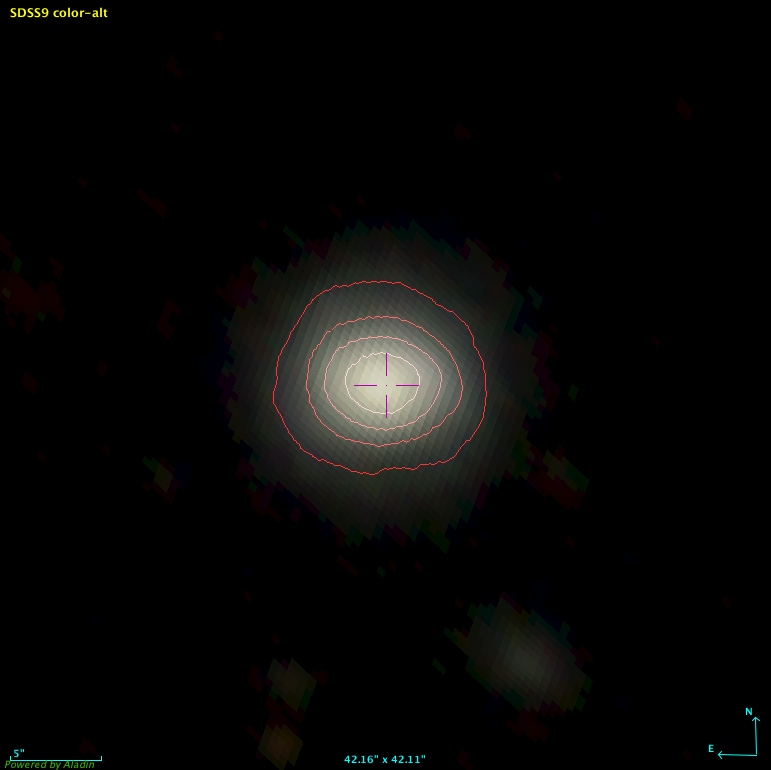}
    \caption{Image cutouts of 10 blue ellipticals ($NUV-r \leq 3$) obtained from the Aladin Sky Atlas tool. Isophote contours, in red, emphasize the smoothness of the light distribution and lack of substructure. All images have a field of approximately $48''$ x $48''$, and are a SDSS9 $g, r, i$ CDS color composition.}
    \label{fig_random10}
\end{figure*}

In Figure \ref{fig_CMD} we show the CMDs for all galaxies with a valid morphological classification, spirals, and ellipticals. Figure \ref{fig_CMD} shows that the typical colour-morphology relation is clearly reproduced, the blue cloud being mostly populated by spirals, and the red sequence, by ellipticals. Blue ellipticals appear to be relatively common. Visual inspection of a random subsample of 10 of these objects (with $NUV-r \leq 3$) using the Aladin Sky Atlas3 (Figure \ref{fig_random10}) shows indeed a lack of galactic structures such as arms or bars, reinforcing the effectiveness of the method in recovering early-type morphologies.


We also include in the CMDs of Figure \ref{fig_CMD} the location of galaxies with GALEX coverage but no UV-detection. In order to do this, we calculate a lower limit for $(NUV-r)_0$ adopting an absolute UV limit magnitude of $M_{UV} = -15$, given that $95\%$ of our sample is brighter than this. Considering that we have redshifts at hand, these undetected objects are likely intrinsically faint in the UV; as can be appreciated in Figure \ref{fig_CMD}, the locations of these upper limits are consistent with objects in the red-sequence, meaning that these galaxies are no longer forming stars. With the exception of a few spirals, the distribution of these lower limits emphasizes that our sample is reasonably complete up to $(NUV-r)_0 \sim 5$. 


We note that for our analysis, we define as members of the blue cloud and of the red sequence those galaxies that lie within $1\sigma$ from the blue and red peaks of the $(NUV-r)_0$ distribution. In order to evaluate whether our results are dependent on this definition, we also display the values of $1.5\sigma$ around these peaks. 



\subsection{SFA \textit{vs.} $(NUV~-~r)_0$}
\label{sub_results_sfa}


To evaluate the dependence of SFA on colour and morphology, we separate our galaxy sample into spirals and ellipticals (see Section \ref{Chapter2}). We divide these subsets into colour bins of $\Delta(NUV~-~r)_0 = 0.5$, and determine the median and distribution of SFA values in each bin for each morphological type. (Table \ref{tab_stats_dl}).



\setlength\tabcolsep{1.5pt}
\begin{table}
    \begin{tabularx}{\columnwidth}{XXXXX}
    \hline
    \hline
        \textbf{spi sample}\textsuperscript{a} & \textbf{ell sample}\textsuperscript{a} & \textbf{SFA$_{spi}$}\textsuperscript{b} &\textbf{SFA$_{ell}$}\textsuperscript{b} & \textbf{pvalue}\textsuperscript{c} \\
    \hline
    \scriptsize{\textit{NUV-r < 1:}} \\
    408 & 59 & -1.24 & -4.86 & 0.00 \\ 
    1458 & 163 & 0.21 & -0.01 & 0.12 \\ 
    \hline
    \scriptsize{\textit{Blue Cloud:}} & & & & \\
    2512 & 265 & 0.48 & 1.23 & 0.00 \\ 
    2666 & 226 & 0.60 & 1.38 & 0.00 \\ 
    2036 & 169 & 0.48 & 1.39 & 0.00 \\ 
    \hline
    \scriptsize{\textit{Green Valley:}} & & & & \\
    1627 & 98 & 0.11 & 1.48 & 0.00 \\ 
    1260 & 109 & -0.30 & 0.96 & 0.01 \\ 
    839 & 139 & -0.33 & 0.29 & 0.00 \\ 
    564 & 144 & -0.24 & 0.87 & 0.00 \\ 
    369 & 210 & 0.46 & 1.15 & 0.00 \\ 
    \hline
    \scriptsize{\textit{Red Sequence:}} & & & & \\
    269 & 432 & 0.88 & 1.43 & 0.00 \\ 
    158 & 493 & 1.60 & 1.69 & 0.07 \\ 
    \hline
    \scriptsize{\textit{NUV-r > 5.5:}} \\
    
    19 & 90 & 1.65 & 2.41 & 0.05 \\
    \hline
    \hline

  \end{tabularx}
  \caption{\textbf{SFA results}: Galaxy sample subdivided into colorbins of $\Delta(NUV-r)_0=0.5$, together with the median $SFA$ values. The last column states the statistical relevance to differentiate the spiral sample and elliptical sample for given colorbin. \textsuperscript{a} \textit{binned sample size}, for ellipticals and spirals; \textsuperscript{b} \textit{binned SFA median}, for ellipticals and spirals; \textsuperscript{c} \textit{Permutation estimated by Monte Carlo} $p-value$.} 
  \label{tab_stats_dl}
\end{table}



\begin{figure*}
    \centering
    \includegraphics[width=15cm]{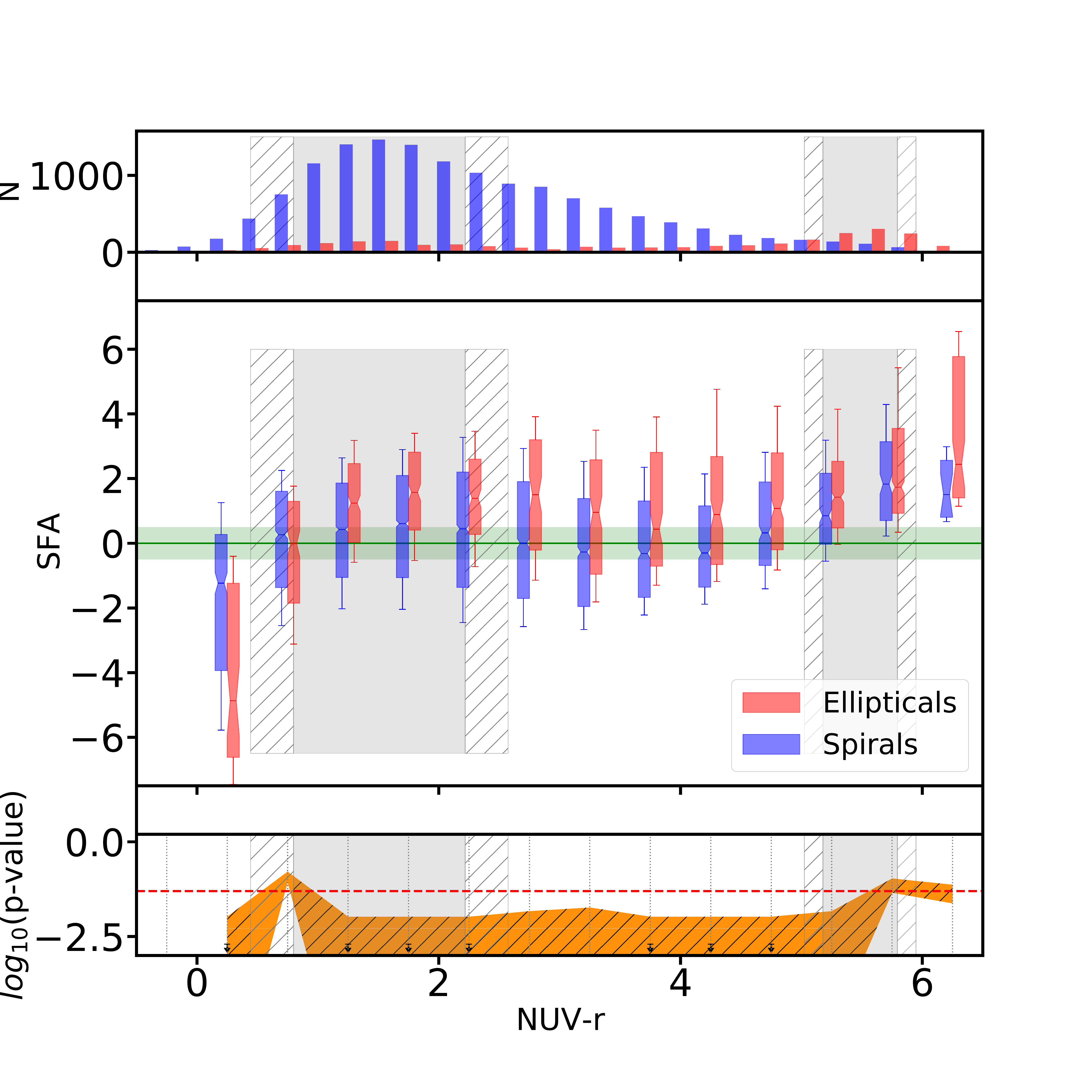}
    \caption{\textbf{Top panel:} Colour-coded histogram of $(NUV~-~r)_0$ distribution for spirals (blue) and ellipticals (red). \textbf{Middle panel:} SFA \textit{vs.} $(NUV~-~r)_0$ trends. Each color bin has two boxplots color-coded for spirals (blue) and ellipticals (red), representing the median SFA value and extending from the $25\%$ to the $75\%$ quartiles. The lower and upper bars encompass between 15$\%$ and 85$\%$ of the distributions, respectively. Each boxplot x-axis position was slightly offset for better display. The shaded green area represents SFA values consistent with 0, considering typical uncertainties of 0.5 (\citealp{martin2017quenching} - Appendix B)}. 
    \textbf{Bottom panel:} \textit{p-value} from the permutation statistical test. The shadowed orange area stands for the range of values for \textit{p-values} with 99$\%$ of confidence. The horizontal dashed red line shows a typical $p-value$ threshold of $0.05$. All three panels display shaded gray areas highlighting the positions of the blue cloud and the red sequence. Adjacent hatched areas represent regions of $1.5~\sigma$ around the distribution peaks in blue and red, showing that the results do not strongly depend on the definition of the green valley. From the SFA trends together with the \textit{p-value} results, we can argue that, in general, elliptical galaxies present higher SFA distributions than spiral galaxies with similar colours.
    
    \label{fig_sfa_morphology}
\end{figure*}

Figure \ref{fig_sfa_morphology} shows the SFA trends as a function of $(NUV~-~r)_0$ colours for our sample, divided by morphology, displayed in boxplots in order to show the distribution of values within each bin. The SFA values at each colorbin for spirals and ellipticals -- shown as blue and red boxplots\footnote{https://matplotlib.org/3.1.1/api/\_as\_gen/matplotlib.pyplot.boxplot.html}, respectively -- show that ellipticals with intermediate colors (i.e., lying within the green valley region, between $2.3\leq NUV-r \leq 5.1$) have higher median SFA values than spirals within the same region. This indicates that ellipticals within the green valley are undergoing faster quenching. We also note that the SFA values for spirals smoothly vary from negative (i.e. \textit{burst}; $NUV~-~r~<1$) to positive values (i.e. \textit{quench}; $NUV~-~r~>4.5$), with some color-bins consistent within uncertainties with $SFA~=~0$ ($-0.5 \leq SFA \leq 0.5$; \citealp{martin2017quenching}). Ellipticals, on the other hand, have a more abrupt SFA change from negative to positive at $NUV~-~r~\sim~1$ followed by a mostly positive SFA trend. In fact, for $NUV-r \geq 1$, $50-75\%$ of ellipticals display SFA values greater or equal to 0 in every colour bin, and at least $50\%$ of ellipticals in all bins show SFA greater than 0.5 (above the uncertainty threshold). This highlights the quenching behaviour for most of these objects. These distinct behaviours suggest different evolutionary paths for spirals and ellipticals. We discuss this further in Section \ref{sub_results_sfa_morphology}. We note that the consistency in behaviour, with ellipticals displaying higher SFA median, lower and upper quartile values than spirals across a wide range of intermediate colors, is a strong indication that the result is significant. 


In order to verify the reliability of our results,  we performed the \textit{Permutation estimated by Monte Carlo} statistical test, using \texttt{R} libraries, to verify whether the two populations -- spirals and ellipticals SFA distributions -- in each color bin are simply subsamples taken from one same parent population. This is achieved by considering the difference in their mean values and determining the probability of recovering this difference if the subsamples are defined randomly.  We first consider the difference $\Delta_{SFA}$ between ellipticals and spirals within each color bin. Starting from the null hypothesis that these samples arise from the same parent population, the probability of recovering the same difference $\Delta_{SFA}$ between the samples should be high if these samples were defined at random. A permutation re-sampling is carried out precisely to verify this. The samples are mixed and separated into two samples of the same size as the original samples; their means are again calculated, as well as the differences between these values. This process is repeated 999 times, so that a distribution of mean differences is obtained. With this resulting distribution, shown in the bottom panel of Figure \ref{fig_sfa_morphology}, we can determine the probability of obtaining the measured difference in our sample.


We note that within each bin a \textit{p-value} of $0.05$ or less allows us to discard the null hypothesis that the SFA values of the two morphological subsets arise from the same parent sample with a $5\%$ chance of error; that is, \textit{p-values} lower than $0.05$ allow us to robustly conclude that the SFA values for the two morphological subsets within the green valley are indeed distinct. The distribution of \textit{p-values} along the $NUV-r$ color sequence shown in Figure \ref{fig_sfa_morphology} shows all \textit{p-values} within the green valley well below $0.05$ (and typically below 0.01). This robustly confirms that green valley ellipticals show higher SFA values than spirals, their star formation quench more sharply.

\section{Discussion \& Conclusions}
\label{Chapter5}




As discussed by \cite{darvish2018quenching}, SFA values cannot be directly translated into \textit{quenching}/\textit{bursting} timescales, since they are not linearly related. We can still interpret the physical meaning of the SFA results qualitatively and comparatively, especially when contrasting two samples, in this case,  that of spiral and elliptical galaxies.


\subsection{SFA \textit{vs.} $(NUV~-~r)_0$: morphology dependency}
\label{sub_results_sfa_morphology}

A scenario where gas supply is disrupted by certain mechanisms (e.g. mass quenching -- \citealp{cattaneo2006modelling}; interruption of cold streams and shock heating -- \citealp{dekel2006galaxy}) would result in a smooth transition through the green valley, with SFA median values close to zero. This is what we see in the case of spirals entering the green valley region from the blue cloud (see Figure \ref{fig_sfa_morphology}). These results are in agreement with those found in \cite{schawinski2014green}, where the authors also suggest that secular processes (i.e, dynamic interactions and instabilities with timescales greater than 1 Gyr) are responsible for quenching spirals that manage to keep their discs as they become red systems. Once the gas supply is blocked, the galaxy slowly evolves through the green valley. Secular processes can also prompt gas depletion: e.g., bars induce gas inflows towards the central regions, triggering SF and accelerating gas consumption. Although the cause-and-effect relation is up for debate (e.g., gas-poor discs also favour the formation of bars \citep{Athanassoula2013}, we do know there is a correlation between quenching and secular processes, since the bar fraction is much higher for red discs \citep{Masters2011}. This can explain the late quenching ($SFA>0$) for spirals beyond $NUV-r \sim 4$. We would like to highlight, however, that this increase in SFA may also be due the fact that towards red colors, our sample is incomplete (see discussion below).

We note that in this scenario, as the galaxies migrate from blue cloud through green valley into the red sequence, they do not alter morphology but rather the cessation of star formation renders the spiral redder. 

\begin{figure}
    \centering
    \includegraphics[width=
    \columnwidth]{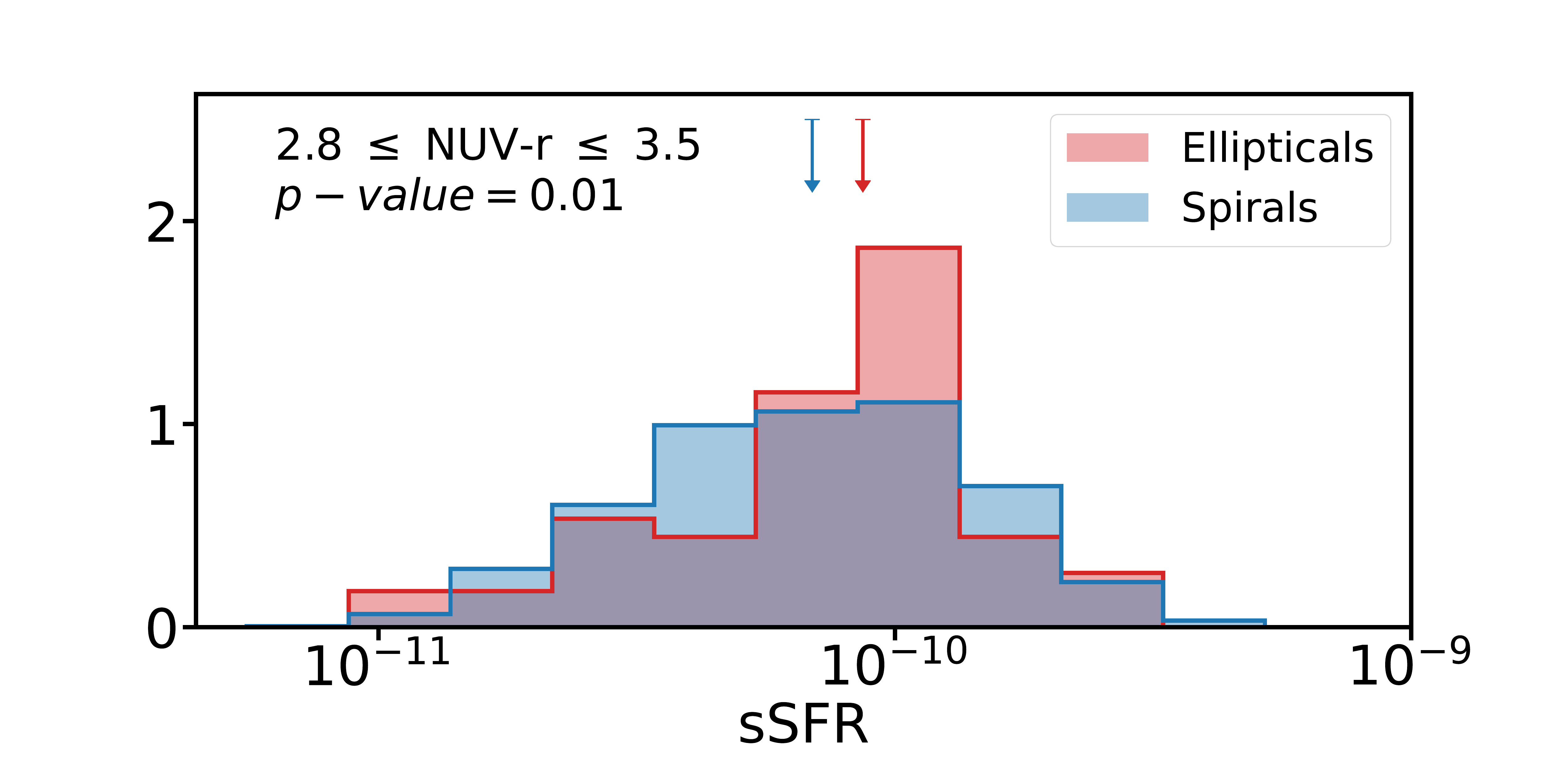}
    \caption{Distribution of sSFR for spirals (blue) and ellipticals (red) for the bluest part of the green valley ($2.8 \leq NUV-r \leq 3.5$). The arrows indicate the median value of the distribution. Entering the green valley, ellipticals have higher values of sSFR, indicating a faster consumption of the gas. The $p-value = 0.01$ confirms the statistical difference between the sample distributions.}
    \label{fig_sSFR_greenvalley}
\end{figure}

The behaviour of elliptical galaxies in Figure \ref{fig_sfa_morphology} is different, with positive and high SFA values for most of the colours beyond $NUV-r = 1$. This suggests an evolutionary pathway of ellipticals through the green valley that involves a mechanism that not only disrupts the gas supply, but also removes, hastily consumes or stabilizes the gas in the galaxy against gravitational collapse (e.g., major merger, AGN feedback, starbursts triggers, morphological quenching). As a consequence, a sudden decrease in SFR as the galaxy enters the green valley results in the higher values measured for the SFA parameter. One could expect, therefore, higher values of sSFR for ellipticals than for spirals in the green valley. Indeed, elipticals on the blue end of the green valley do show higher sSFR\footnote{https://www.sdss.org/dr14/spectro/galaxy\_mpajhu/} than spirals (Figure \ref{fig_sSFR_greenvalley}), based on the SFR and M* measurements by \citet{kauffmann2003stellar}, \citet{brinchmann2004physical}, and \citet{salim2007uv}, indicating a more violent transition between star-forming and passively evolving for early-type galaxies.

This behaviour supports the scenario in which elliptical galaxies experience quenching processes with shorter timescales than spirals. Previous works have reached similar conclusions. \cite{schawinski2014green}, for example, used an exponential-decay SFH for green valley galaxies in the local universe and obtained the same results. The authors conclude that a short episode of enhanced star formation in these systems, followed by rapid quenching, was likely associated to merger activity. At intermediate redshifts, work by \cite{nogueira2018star}, reached a similar conclusion, based on spectroscopic data and stellar population analyses of $z \sim 1$ galaxies with reliable morphological classification.


Recent results show that, although gas-rich mergers may be the main physical mechanism to change morphological structure in galaxies while simultaneously affecting star formation, secondary mechanisms may also be needed (e.g., \citealp{smethurst2016galaxy}, \citealp{rodriguez2017role}). These would not only quench the star formation, but also prevent the galaxy from forming stars again, such as AGN feedback triggered by merger (e.g. \citealp{sanders1990ultraluminous}; \citealp{hopkins2006relation};  \citealp{croton2006many}; \citealp{schawinski2009moderate}; \citealp{fabian2012observational};  \citealp{rodriguez2017role}). In addition, \cite{dubois2016} showed through simulations that, in order to build massive ellipticals, AGN feedback is necessary to prevent the final object from rebuilding a disk, guaranteeing the establishment of an early-type morphology.


\cite{martin2017quenching} explore the relation between SFA and AGN and find that at intermediate specific SFRs ($10^{-3}-10^{-1}$ yr$^{-1}$), galaxies from the SDSS database (\citealp{martin2007uv}) present higher values of SFA by a factor of $2-3$. This suggests that AGN may indeed contribute to faster quenching in galaxies. However, as the authors themselves emphasise, the presence of AGN is correlated/intertwined with many other galaxy properties. So, although investigating AGN signatures may contribute greatly to evolutionary picture we have painted by analysing SFA and galaxy morphologies, a full picture requires a more refined analysis of galaxy morphologies. Our results support the idea of a morphological transformation caused by  major/minor mergers in tandem with intense quenching from violent interactions and further analysis on AGN dependence are beyond the scope of this work.

Since we use a non-parametric methodology, it is also possible to analyze the SFA distribution not only for the green valley region, but for the entire $NUV-r$ range. This allows us to, for the first time, argue that star formation histories of different morphological types are distinguishable not only within the green valley region. We find that elliptical and spiral galaxies in the blue cloud already show distinguishable star formation histories: while spiral galaxies have an approximately constant SFR -- reflected in SFA values closer to $0$ -- elliptical galaxies in the blue cloud already exhibit a \textit{quenching} behaviour, with positive and high SFA values. This result is confirmed by very low \textit{p-value}, confirming that SFA values for these two populations are statistically distinct even within the blue cloud, at $1.0 \lesssim NUV-r \lesssim 2.3$. One possibility to explain this behavior is that these blue ellipticals are passing through the blue cloud region for the first time, after a major merger event. As demonstrated by \cite{mendez2011aegis}, the largest merger fractions within the CMD diagram are located in bluer colors. Furthermore, simulations show that, after major mergers, galaxies go through a change in their morphology prior to suffering a drop in SFR (e.g. \citealp{dubois2016}; \citealp{martin2018role} \citealp{tachella2019}; \citealp{joshi2020fate}), which explains why we find elliptical galaxies quenching within the entire $NUV-r$ range. Preliminary results by Sampaio et al. (in prep.) also suggest a similar behaviour in galaxy clusters. They find that brighter galaxies change their T-Type morphologies more rapidly, while fainter satellites decrease their star formation rates first. This is probably as a result of environmental processes that do not cause any abrupt morphological transformations in satellite galaxies, such as ram-pressure stripping or strangulation. Since the sample in this work is limited to objects brighter than $M_r \sim -20.40$, we cannot at this point verify the findings by Sampaio et al. In this sense, future work is required to probe SFA values for fainter, less massive galaxies.


Galaxies also undergo rejuvenation processes, usually through an encounter with a gas-rich companion and subsequent episode of star formation (e.g. \citealp{rampazzo2007galaxy}; \citealp{martin2017quenching}; \citealp{pandya2017nature}). Some ellipticals could reach bluer colours in this manner, and particularly those with $NUV-r \lesssim 1$, $SFA < 0$, are indeed bursting. Nevertheless, ellipticals redder than $NUV-r \gtrsim 2.3$ are typically quenching, regardless of the formation process prior to that stage.


The SFA behaviour of galaxies within the red sequence does not seem to depend on galaxy morphology as strongly: SFA values for spirals and ellipticals (Figure \ref{fig_sfa_morphology}, middle panel) are statistically indistinguishable, as opposed to bluer colours (Figure \ref{fig_sfa_morphology}, lower panel). These results indicate that the main differences for the evolutionary paths between spirals and ellipticals probably are found at the beginning of the quenching process and, once the galaxy becomes passive, there is less significant dependence on the responsible physical mechanism. However, we remind the reader that, as discussed in Section \ref{sub_sec_galaxysample}, our sample is not complete for the less massive and redder galaxies. Furthermore, one would expect that $SFA \approx 0$ for red sequence galaxies, which would already be quenched. Indeed \citeauthor{martin2017quenching} show that galaxies reach $SFA\approx0$ beyond $NUV-i > 4$. Nevertheless, since our sample is not complete for such colours, the bias introduced by the need for UV detection (Section \ref{chp4}) is potentially skewing our measurements towards objects with remaining traces of star formation activity. Deeper UV and spectroscopic data are required to properly measure the SFA of red/faint galaxies

\subsubsection{Quenching timescales and physical processes}

\begin{figure}[!h]
        \centering
        \includegraphics[width=\columnwidth]{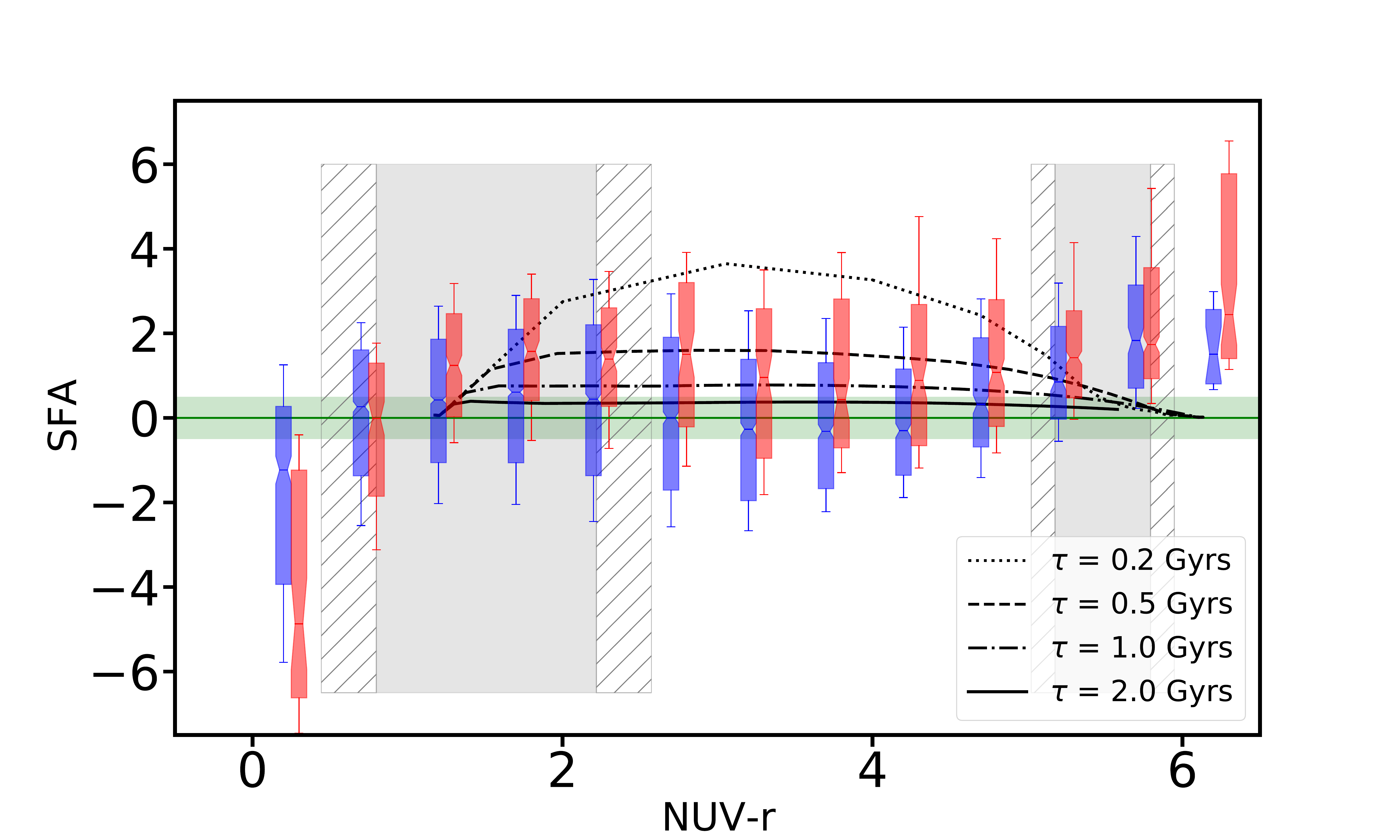}
        \caption{Same as Figure \ref{fig_sfa_morphology}, with models of e-folding quenching SFH (\citealp{bruzual2003stellar}) with different timescales: 2 Gyrs (solid line), 1 Gyr (dot-dashed line), 0.5 Gyrs (dashed line) and 0.2 Gyrs (dotted line). }
        \label{fig_BC03}
\end{figure}

One should be careful when comparing SFA values with quenching timescales, given that the relation is not univocal. Nonetheless, in order to allow for a quantitative comparison with previous results in the literature, we investigated SFA behaviour for synthetic stellar population models with different e-folding times (\citealp{bruzual2003stellar} - hereafter BC03) with timescales of $0.2$, $0.5$, $1.0$, and $2.0$ Gyr (Figure \ref{fig_BC03}). We consider constant SFR for $6$ Gyr followed by different an exponential decline in SFR as ${\rm SFR} = {\rm SFR}_0\;{\rm exp}(-t/\tau)$, assuming a Chabrier IMF and solar metallicity. We follow the SFA definition from \cite{martin2007uv} and measured it from the stellar population models as following:

\begin{equation}
    SFA = \frac{NUV-i (t=t_0+0.3\;\text{Gyr}) - NUV-i (t=t_0)}{0.3\;\text{Gyr}},
\end{equation}

\noindent with $t_0$ at each timestep.

As Figure \ref{fig_BC03} shows, the median SFA values for spirals in the green valley are consistent with quenching timescales $\tau \sim 1- 2$ Gyr, with the upper quartiles of most bins within $\tau \geq 0.5$ Gyr. On the other hand, for ellipticals in the same region, we have median SFA values with $\tau \leq 1$ Gyr, with the upper quartiles reaching values of $0.2 Gyr \lesssim \tau \lesssim 0.5$ Gyrs.

\cite{martin2007uv} have noted that objects with fast quenching timescales are less likely to be observed in the green valley, since they cross the region faster. With that in mind, we emphasize that a distribution corrected for this detectability bias would show a significant number of objects with $\tau \lesssim 0.5$ Gyr.


The values determined in this work ($\tau_{spi} \gtrsim 1$ and $\tau_{ell} \sim 0.5$ Gyr -- correspond to the expected timescales for fast and slow quenching processes, respectively (e.g. \citealp{schawinski2007effect}; \citealp{gonccalves2012quenching}; \citealp{martin2017quenching}). In other words, spirals show quenching timescales that correspond to processes where gas supply is disrupted (e.g. mass quenching -- \citealp{cattaneo2006modelling}; interruption of cold streams and schock heating \citealp{dekel2006galaxy}; secular procceses, dynamic interactions and instabilities -- \citealp{schawinski2014green}), while ellipticals have quenching timescales that correspond to processes where the gas reservoir is either disrupted, hastily consumed or stabilized against gravitational colapse (e.g. major mergers -- \citealp{schawinski2014green}, \citealp{nogueira2018star}; AGN feedback -- \citealp{fabian2012observational}; minor mergers and morphological quenching -- \citealp{martig2009morphological}, \citealp{brennan2015quenching}, \citealp{martin2018role}).


\cite{joshi2020fate} used the IllustrisTNG simulations to show that galaxies that were disc-like and are non-disc in $z=0$ first undergo morphological change, followed by a drop in SFR. This is in good agreement with our results, including the presence of blue star-forming ellipticals for colors $NUV-r \leq 1.0 $. These galaxies would be the result of a violent event that have recently undergone a substantial structural transformation, and will rapidly lose their gas reservoir, moving into the red sequence at a fast pace.


We should note that BC03 models reach the red sequence with $SFA \approx 0$, specially for $NUV-r\geq 5$, as expected. As previously discussed, this sample is not complete at the red end of the colour-magnitude diagram. This could result in a selection bias towards galaxies with $UV$ emission, i.e. objects that are not yet fully quenched.

\subsection{SFA \textit{vs.} $(NUV~-~r)_0$: mass dependence}
\label{sub_results_sfa_mass}

The works by \cite{martin2017quenching} and \cite{darvish2018quenching} have emphasized how SFA depends strongly on the stellar mass of galaxies. Therefore, in this section we investigate the impact stellar mass may have on our analysis of SFA as a function of galaxy morphology.

\begin{figure*}
    \centering
    \includegraphics[width=\columnwidth]{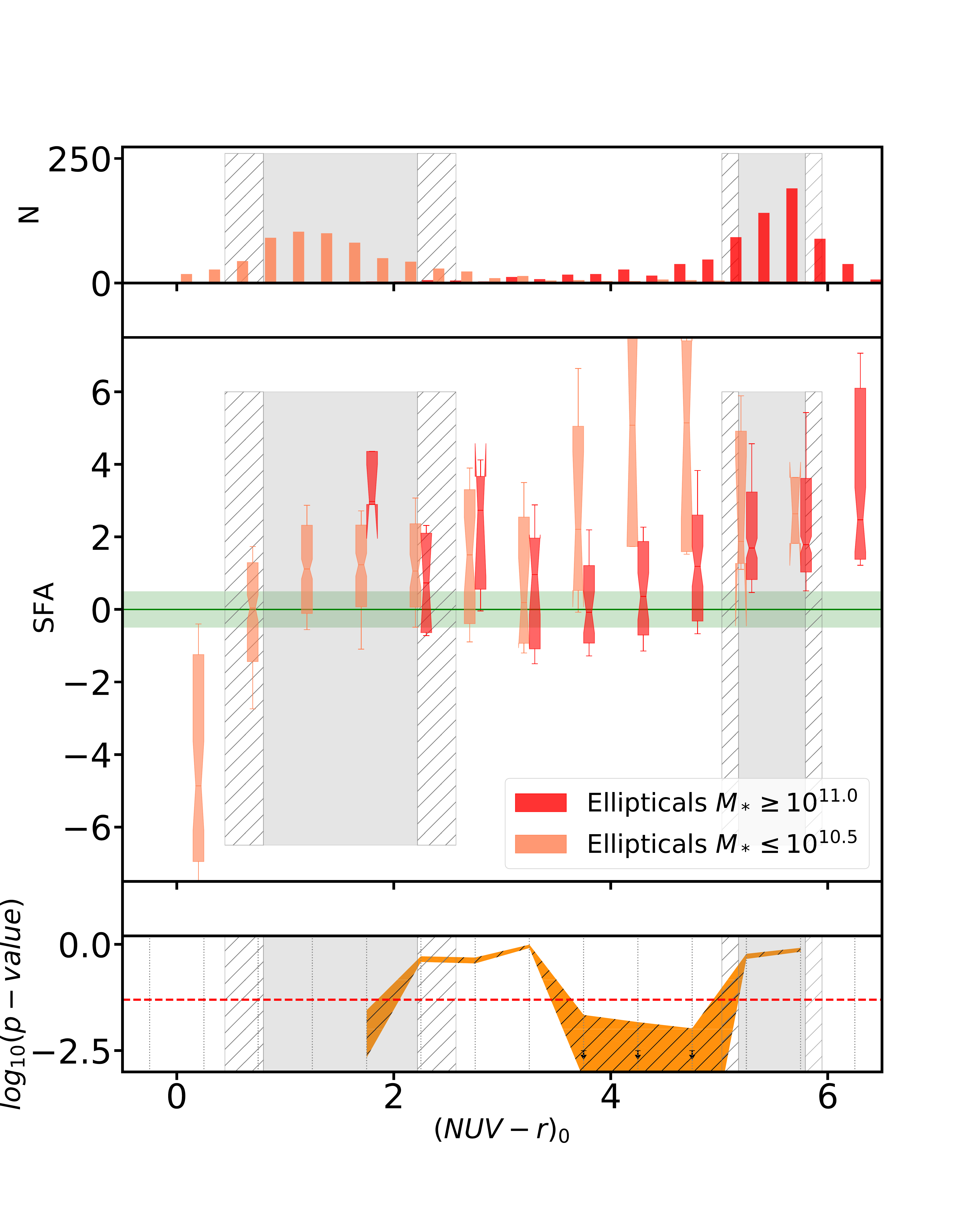}
    \includegraphics[width=\columnwidth]{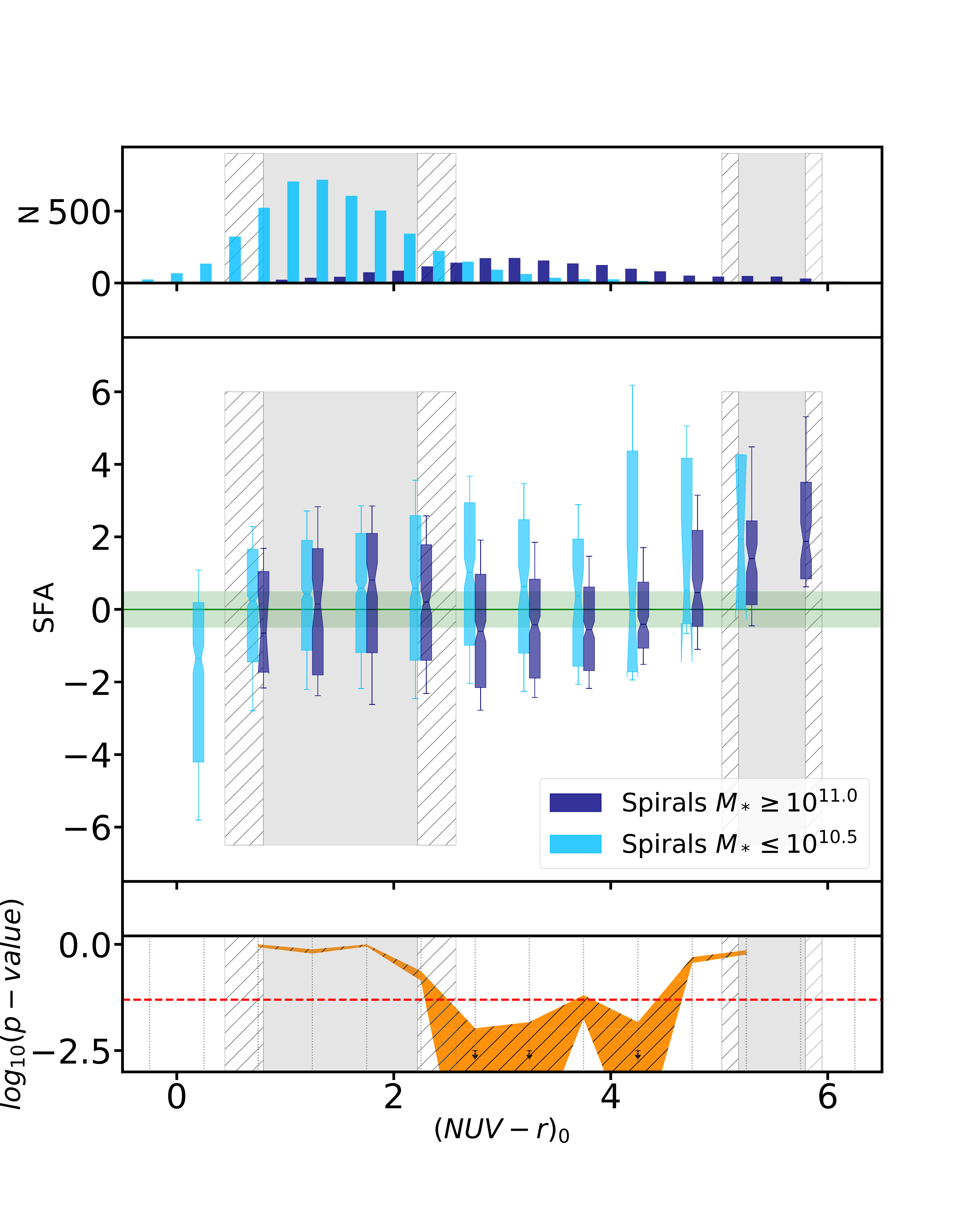}
    \caption{\textbf{Top} panels: $(NUV-r)_0$ distributions for ellipticals (\textbf{left}) and spirals (\textbf{right)} for different mass cuts. \textbf{Middle}: SFA\textit{ vs.} $(NUV - r)_0$ distributions of boxplots for each color bin, representing the median SFA values, the upper and lower quartis and the range of the distribution. In the \textbf{bottom} panel, we show the \textit{p-value} result from permutation, following the analysis in Figure \ref{fig_sfa_morphology}. Spirals proceed through different evolutionary pathways for the green valley, depending on mass, with massive spirals typically bursting. Ellipticals, on the other hand, are all mostly quenching, albeit with different timescales.} 
    \label{fig_sfa_mass}
\end{figure*}

To that end we use the MPA-JHU mass catalogs\footnote[4]{https://www.sdss.org/dr14/spectro/galaxy\_mpajhu/}, based on the works by \cite{kauffmann2003stellar} and \cite{salim2007uv}. The results can be found in Figure \ref{fig_sfa_mass}, where we analyse the SFA behaviour across the CMD in two stellar mass bins:  $M_\star \geq 10^{11.0}$ and $M_\star \leq 10^{10.5}$ solar masses, for the two morphological types considered (ellipticals and spirals). Although our low-mass limit is greater than that in \cite{martin2017quenching} (log $M_\star/M_\odot \sim 10.0$ vs log $M_\star/M_\odot \sim 9.5$), we still cover a wide enough dynamic range in stellar mass to detect a trend of SFA as a function of mass. 

We find that the behaviour of SFA as a function of colour in different stellar mass bins (log $M_\star/M_\odot \geq 11$ and log $M_\star/M_\odot \leq 10.5$) is distinguishable only in the green valley region (Figure \ref{fig_sfa_mass}, bottom panels), for both morphological types. This may be a result of small-number statistics, however, since there are few blue massive and red low-mass ellipticals in our sample. We confirm the different behaviours with the same statistical tests used in Section \ref{sub_results_sfa_morphology} (bottom panels of Figure \ref{fig_sfa_mass}).

We determine the fraction of galaxies quenching and bursting in the green valley region for each morphology. 32$\%$ of low-mass spirals are bursting and 54$\%$ are quenching, beyond the threshold consistent constant star formation rates ($-0.5 \leq SFA \leq 0.5$). Comparatively, for high-mass spirals, we have 45$\%$ and 33$\%$, respectively. Since most green-valley spirals are more massive (top-right panel of Figure\ref{fig_sfa_mass}), we conclude that most of the bursting activity in the green valley happens in high-mass spirals. On the other hand, 21\% of low-mass ellipticals are and 66$\%$ are quenching. Lastly, the figures for high-mass ellipticals are 25$\%$ and 56$\%$, respectively. We conclude that most ellipticals are quenching, regardless of mass. However, median SFA values are higher for less massive ellipticals, meaning these objects display shorter quenching timescales.


These results are in line with \cite{martin2017quenching} and \cite{darvish2018quenching}, who found that massive galaxies are more likely to undergo \textit{bursts} of star formation in the green valley, while less massive galaxies suffer \textit{quenching}. This is consistent with the environmental quenching scenario in which satellite galaxies may lose their molecular gas content to respective central galaxies, triggering a burst in the latter and quenching the former.

The behaviour of galaxies in the red sequence and beyond, however, is dominated by massive galaxies, due to a lack of low-mass, very red objects in our sample. Rejuvenation episodes notwithstanding, the SFA analysis of \cite{darvish2018quenching}, focused on the impact of galaxy environment, points to significant \textit{quenching} in galaxies in dense environments, such as clusters. Taking this into account, along with the correlation between morphology and environment (\citealp{dressler1980galaxy}), where elliptical galaxies are more numerous in dense environments, we understand the quenching behaviour in massive ellipticals as a likely environmental effect, in agreement with our results for the reddest ($NUV-r > 5$) elliptical galaxies. Nevertheless, this might be a result of the UV incompleteness in our sample, and deeper UV data would be needed to confirm these results.

Our results for low-mass ellipticals, on the other hand, are severely hampered by small-number statistics, especially at the red end of the green valley. Qualitatively, however, they also agree with \cite{darvish2018quenching}, who argue that low-mass ellipticals are more subject to environmental effects and typically quenching at a faster rate (higher SFA values). Nevertheless, a much larger sample would be needed to confirm this quantitatively.

We would also like to point out to the reader that SFA median values are more robust for bigger samples. In the green valley, specially for low mass ellipticals, each bin contains only $\sim$10 objects. Considering an individual error in the SFA measurement of $\sigma \approx 1.5$ there is still tentative evidence within the uncertainties, however, for faster quenching in low-mass elliptical galaxies in the green valley.



\subsection{SFA \textit{vs.} $(NUV~-~r)_0$: asymmetry dependence}
\label{sub_results_sfa_assy}

\begin{figure*}
    \centering
    \includegraphics[width=\columnwidth]{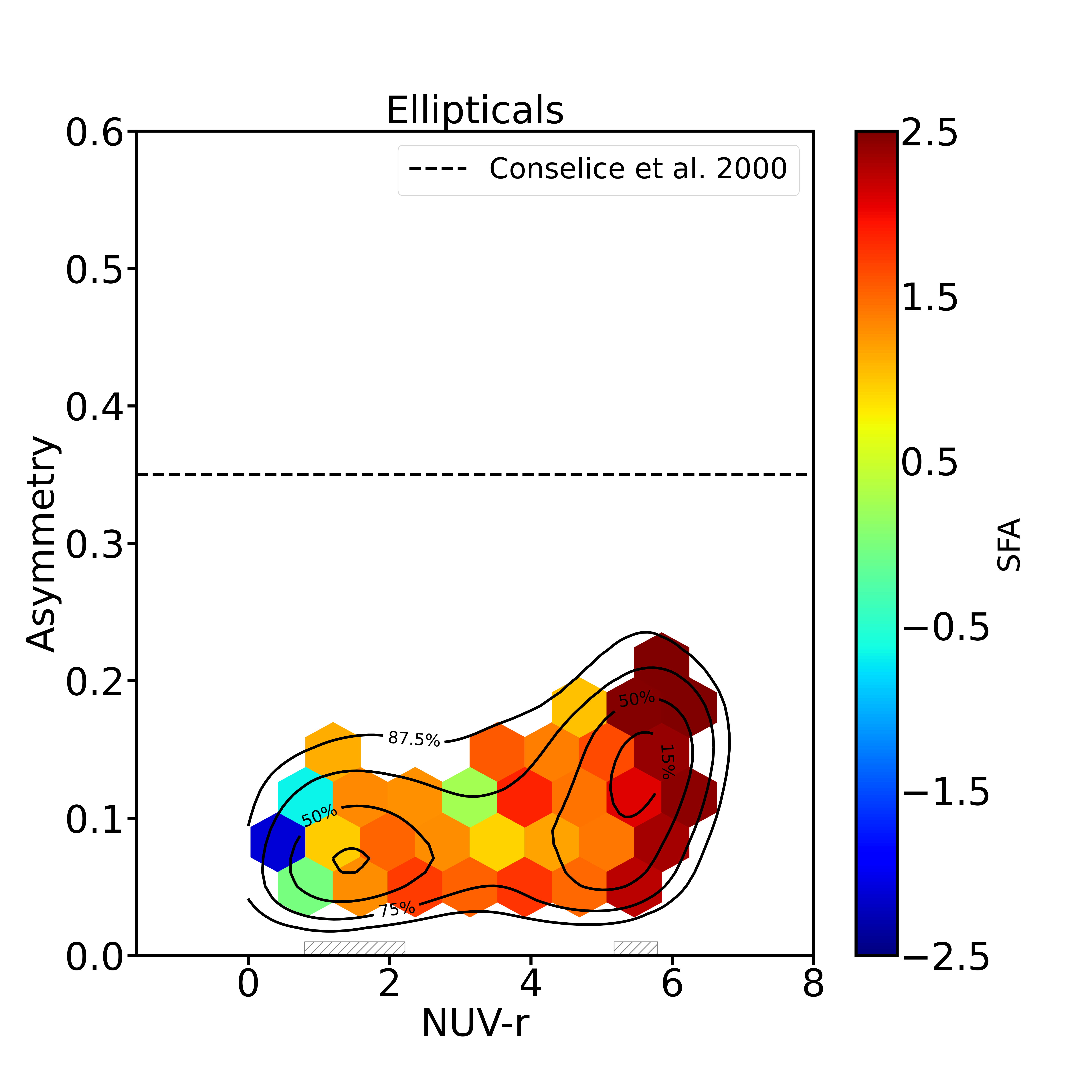}
    \includegraphics[width=\columnwidth]{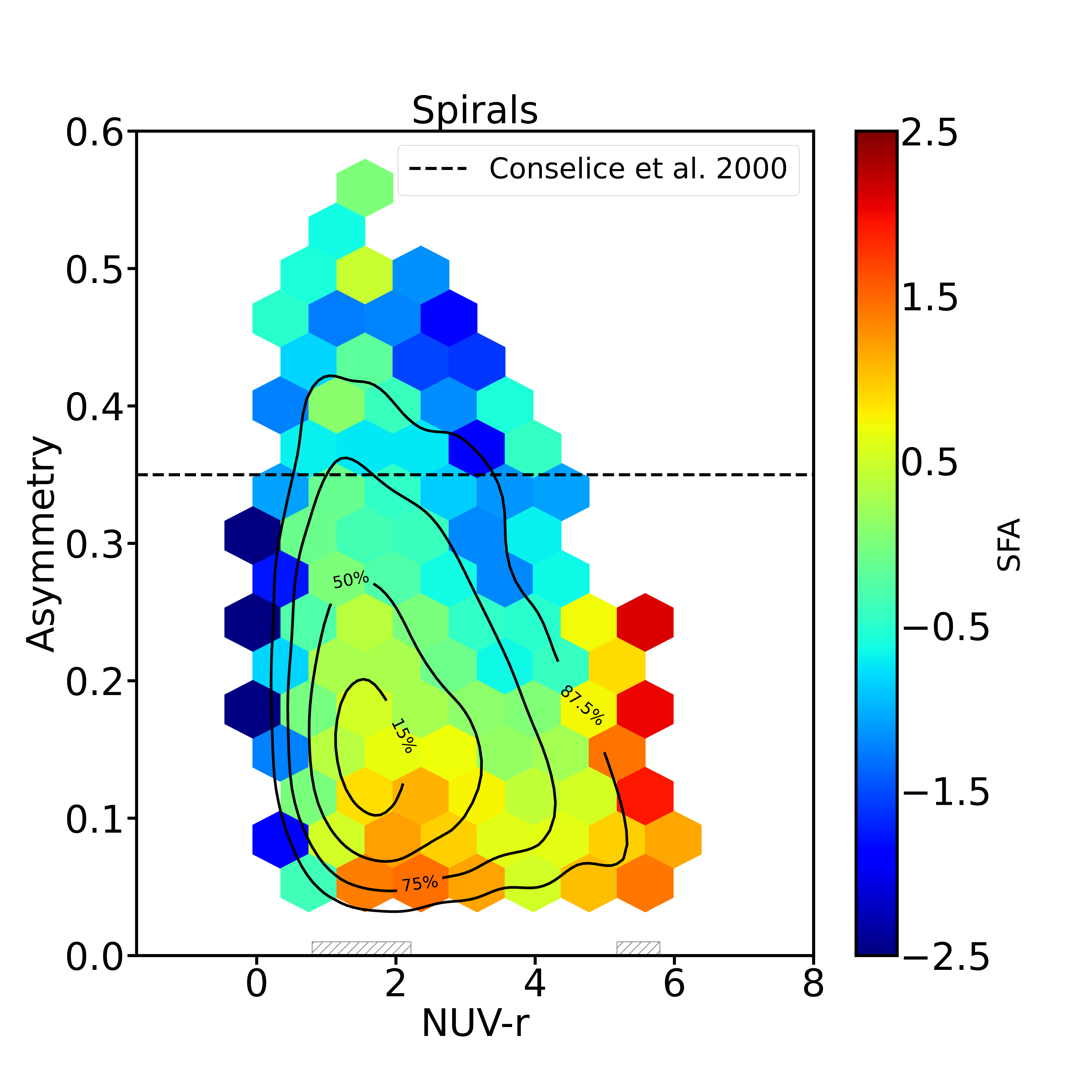}
    \caption{Asymmetry \textit{vs.} $NUV~-~r$ for ellipticals (left) and spirals (right), color-coded with the SFA average bin value for bins with more than 10 galaxies each ($n \geq 10$). The black contours follow from previous Figures, indicating sample percentiles, within each level (15$\%$,50$\%$, 75$\%$ and 87.5$\%$). The hatched gray areas in the bottom indicate the blue cloud and the red sequence, as defined in this work. Lastly, the horizontal dashed line refers to the value stipulated by \protect\cite{Conselice_2000} above which asymmetry values are usually attributed to mergers. There is a clear transition for spirals between $1 \leq (NUV~-~r) \leq 4$, where asymmetric objects are bursting and symmetric ones are passive and quenching. For ellipticals, on the other hand, asymmetry values are always small and SFA values point to \textit{quenching} behavior throughout, with the exception of $NUV-r \leq 1$, which can be explained by post merger objects and rejuvenation processes. These results are in agreement with \protect\cite{morell2020classification}.} 
    \label{fig_sfa_asy}
\end{figure*}

Although major mergers are the usual culprits associated with the morphological transformation of galaxies as they travel across the CMD (e.g., \citealp{di2005energy}), minor mergers also play an important role (e.g. \citealp{martin2018role}). As previously discussed, minor mergers are expected to happen more frequently than major mergers (e.g. \citealp{lotz2011major}) and can enhance star formation and perturbations such as morphological asymmetries (e.g., \citealp{ruiz2020recurrent}). In order to investigate the role of these processes in our sample, we analyse the dependence of SFA on asymmetry for elliptical and spiral morphological types.


The asymmetry value from the Deep Learning catalogue by \citealp{barchi2019machine} was calculated with \texttt{CyMorph} (\citealp{rosa2018gradient}), defined as follows:

\begin{itemize}
    \item \texttt{Asymmetry}: $ A = 1 - s (I ^ {0}, I ^ {\pi}) $, where $ I ^ {0} $ and $ I ^ {\pi} $ are the original and rotated image by 180\textdegree. The $ s()$ function is the \textit{Spearman} correlation function.
\end{itemize} {}

We compute average SFA values in two-dimensional bins within an asymmetry \textit{vs.} $NUV-r$ diagram. The results can be seen in Figure \ref{fig_sfa_asy}, color-coded as a function of SFA average value ($SFA \leq -0.5$ in blue indicating \textit{bursting}, and $SFA \geq 0.5$ in red indicating \textit{quenching} and $-0.5 \leq SFA \leq 0.5$ in green indicating a nearly constant SFR for the past 300 Myrs). The elliptical galaxies display a small dynamic range of asymmetry values ($A \lesssim 0.2$) and no strong dependence of SFA on asymmetry. Elliptical galaxies of all colours are typically quenching, with the exception of very blue objects ($NUV-r \leq 1$), which are bursting. This subsample can be explained by rejuvenation processes or post mergers objects, as discussed in Section \ref{sub_results_sfa_morphology}. Spiral galaxies, on the other hand, display a more complex behaviour as a function of asymmetry. Objects bluer than $NUV-r \leq 1$ are typically bursting and objects redder than $NUV-r \gtrsim 4$ are typically quenching, with no apparent asymmetry dependence. Galaxies within the blue cloud and in intermediate colours, however, depend strongly on asymmetry: asymmetric galaxies are \textit{bursting}, while symmetric ones are either \textit{quenching} or with constant SFR for the last 300 Myrs (green colors), with a threshold distinguishing both behaviours at $A \approx 0.2$.

We would like to draw the reader's attention to the fact that the number of objects in each bin is not constant, as highlighted by the contour levels in Figure \ref{fig_sfa_asy}. Therefore, although there is a strong correlation between asymmetry and SFA, the bulk of galaxies in the green valley show either positive or close to zero SFA values (quenching).

The negative SFA values associated with higher degrees of asymmetry in spirals is in line with the scenario in which (minor) mergers can trigger star formation and cause perturbations in the morphology (e.g. \citealp{kennicutt1987effects}; \citealp{martin2017quenching}; \citealp{ruiz2020recurrent}). This connection becomes particularly significant when considering that an asymmetry value of $A \geq 0.35$ as a reference to identify mergers and interactions (horizontal dashed line in Figure \ref{fig_sfa_asy} -- \citealp{Conselice_2000}, despite slight differences between the methods of quantifying asymmetry). It also agrees with our mass dependence analysis (Section \ref{sub_results_sfa_mass}), according to which massive spiral galaxies consume less massive galaxies, through the process of cannibalism (\citealp{martin2017quenching}), enhancing their star formation and suffering disturbances, which increases the asymmetry index. Indeed, $73\%$ of asymmetric galaxies ($A \geq 0.2$) in the green valley are massive spirals ($M_\star \geq 10^{11} M_\odot$), and $\approx 51\%$ of massive spirals in the green valley have $A \geq 0.2$. 

Furthermore, this result suggests that spirals can evolve through one of two main pathways: a slow, passive evolution or a rapid, more dramatic one. The former group evolves under the effect of secular processes (e.g. \citealp{kormendy2004secular}; \citealp{sheth2005secular};  \citealp{fang2013link};  \citealp{schawinski2014green}; \citealp{bluck2014bulge}; \citealp{nogueira2018star}), gradually exhausting the gas reservoir but maintaining their discs as galaxies reach the red sequence, without going through significant morphological changes. The latter group, on the other hand, experiences mergers and interactions (e.g. \citealp{schawinski2007effect}; \citealp{nogueira2018star}; \citealp{martin2017quenching}), which can enhance star formation, hastily consuming the galaxy's gas reservoir, and provoking the observed asymmetries (e.g. \citealp{ruiz2020recurrent}). These galaxies will then either undergo dynamic cooling or suffer a major morphological change, giving rise to the population of symmetric objects that dominate the red sequence (see Figure \ref{fig_sfa_asy}). 

We also emphasize that symmetric ($A<0.2$) spirals display positive SFA values even in the blue cloud, i.e. even isolated spirals in our sample already display \textit{quenching} behaviour in the blue cloud. Taking into account that our sample is limited to galaxies brighter than $M_{r} = 20.39$ and typically more massive than $M_\star \gtrsim 10^{10}$, this result is in agreement with \cite{peng2010mass}, who found that for low-density environments the quenched fraction is strongly dependent on mass. This could then be explained by mass quenching, where a fraction of galaxies with masses , even in the field, already display a decrease in star formation rate. 

\begin{figure}
        \centering
        \includegraphics[width=\columnwidth]{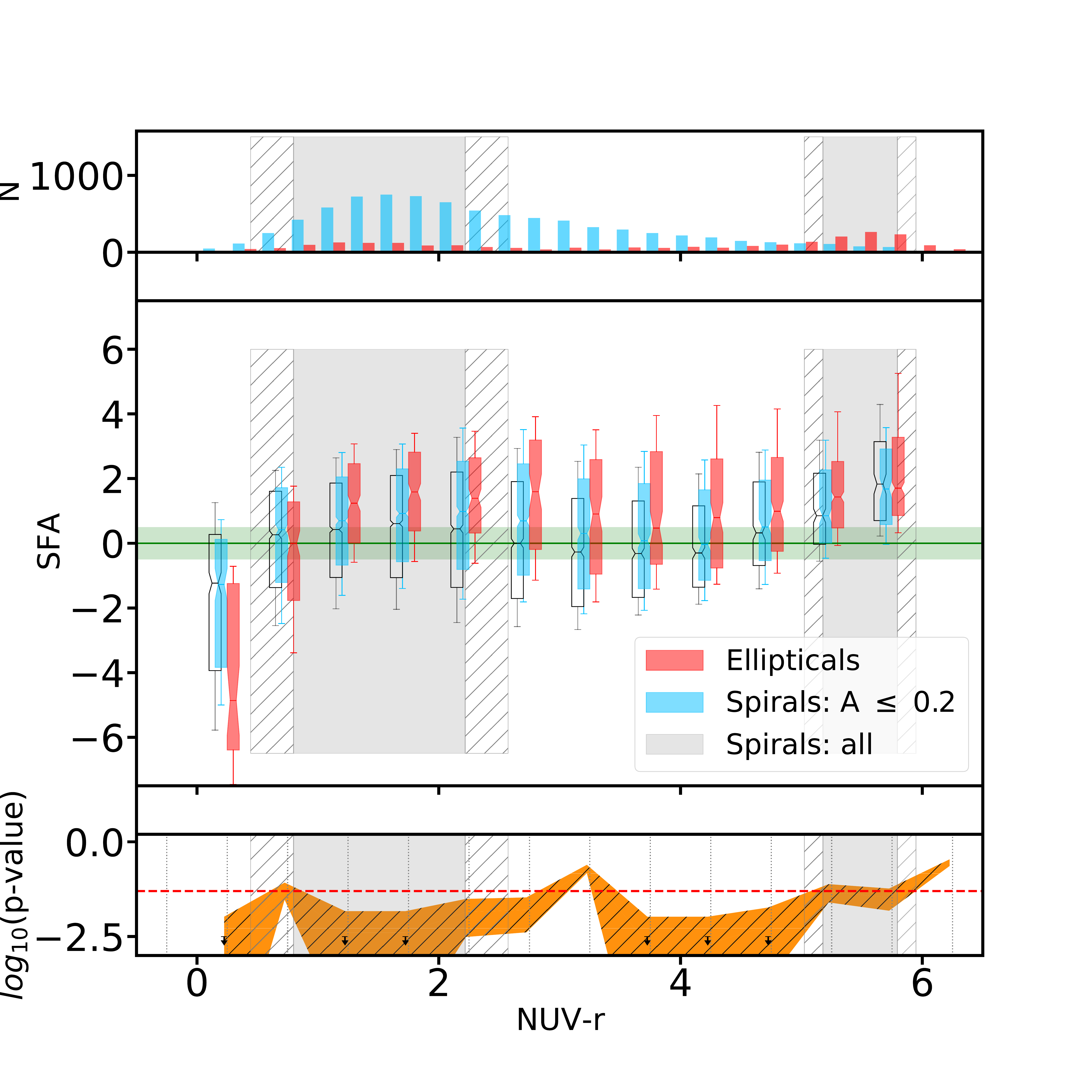}
        \caption{Same as Figure \ref{fig_sfa_morphology}, comparing ellipticals with symmetric spirals ($A < 0.2$). For comparison, we also display the SFA trend for all spirals (Figure \ref{fig_sfa_morphology}) in black boxplots. The comparison between the whole sample of spirals and only symmetric spirals highlight how the latter have higher median SFA values for $1.5 < NUV-r < 4.5$. In the \textbf{bottom panel} we display the \textit{p-values} comparing ellipticals and symmetric spirals only ($A\leq0.2$). From the permutation test, we can affirm that ellipticals and symmetric spirals are statistically distinguishable (with the exception of one bin in $ 3.0 < NUV-r < 3.5$).}
        \label{fig_fig4_sim}
\end{figure}

We also analysed the most symmetric objects separately, selecting only galaxies with $A < 0.2 $ (Figure \ref{fig_fig4_sim}). This cut includes all ellipticals and only the non-disturbed spirals. 

As can be appreciated from Figure \ref{fig_fig4_sim}, the SFA values for spirals are higher among the symmetric galaxies; this is to be  expected from Figure \ref{fig_sfa_asy}. However, we emphasize that \textit{there is still some difference between the SFA trends for symmetric spirals and ellipticals, even when we are dealing mostly with quenching galaxies} (confirmed by \textit{p-values} below $0.05$, with the exception of the color-bin of $NUV-r \sim 3.5$). This supports the scenario where disc galaxies (that become red spirals) and ellipticals have different evolutionary pathways for quenching, where the first will evolve through secular processes and maintain its disc (e.g. \citealp{kormendy1982observations}; \citealp{sheth2005secular}; \citealp{fang2013link};  \citealp{schawinski2014green}; \citealp{bluck2014bulge}; \citealp{nogueira2018star}), and the second is the consequence of interactions and major/minor mergers (e.g. \citealp{springel2005simulations};  \citealp{schawinski2007effect}; \citealp{martig2009morphological}; \citealp{martin2017quenching}; \citealp{nogueira2018star}; \citealp{martin2018role}).

\begin{figure*}
    \centering
    \includegraphics[width=1.6\columnwidth]{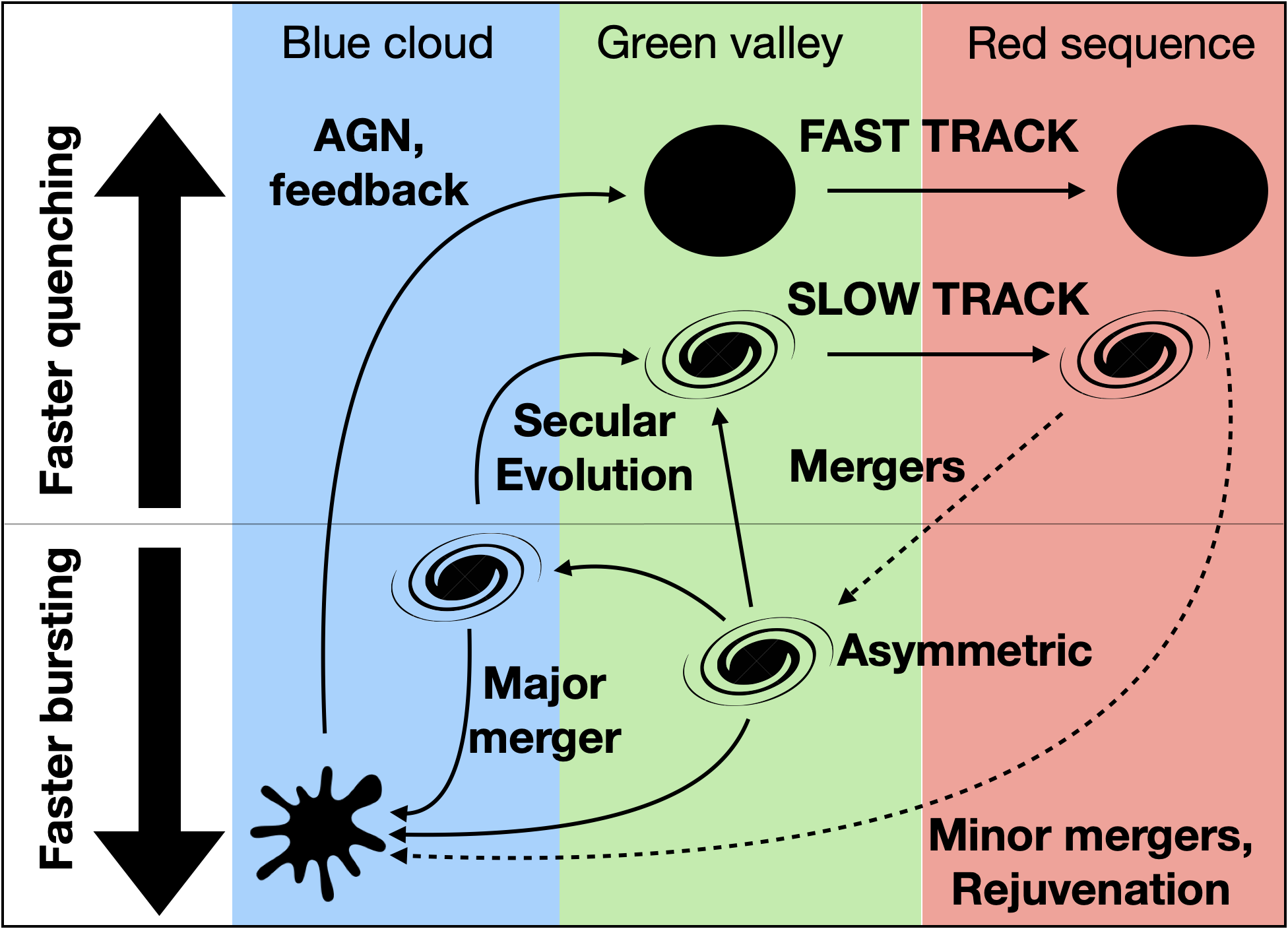}
    \caption{Outline of processes discussed in this work. Using the same axes as Figure \ref{fig_sfa_morphology} (SFA \textit{vs.} $NUV-r$), we present the complexity of evolutionary pathways of spiral and elliptical galaxies across the CMD. The schematic is color-coded for \textit{Blue Cloud/Starburst}, \textit{Green Valley} and \textit{Red Sequence} regions. In summary, secular evolution that does not affect spiral morphologies represent a slow track through the green valley, while violent transformations resulting from mergers and subsequent AGN feedback move a galaxy faster through to the red sequence. Rejuvenation might also occur, and asymmetric spirals are mostly undergoing a bursting event. Massive ellipticals might also undergo temporary rejuvenation, by cannibalizing gas-rich objects infalling to dense environments.}
    \label{fig_CSF21}
\end{figure*}

Lastly, we summarize our findings in a schematic view as shown in Figure \ref{fig_CSF21}, representing physical processes responsible for changing galaxy colours within the SFA \textit{vs.} $NUV-r$ diagram (as in Figure \ref{fig_sfa_morphology}). We emphasize the diversity of possible evolutionary pathways: in the green valley, there are elliptical galaxies with higher SFA values than spirals, indicating the former are in a ``fast-track'' to the red sequence, as opposed to the ``slow-track'' for the latter. At the same time, also within the green valley, there is also a bursting population, composed mostly of asymmetric massive spirals, resulting of interactions and minor mergers of at least one object with an older stellar population. Blueward of the green valley, there are major mergers between spirals forming quenched ellipticals, as well as blue ellipticals undergoing brief rejuvenation episodes (which might be intrinsically bluer than the peak of the distribution within the blue cloud, characterizing a starbursting population). Finally, the red sequence is composed of quenched ellipticals (mostly) and red spirals, which will remain there unless there is an encounter with a gas-rich object.


\section{Summary}
\label{Summary}


We study the passage of galaxies across the color magnitude diagram with the aim of understanding the different evolutionary trajectories that galaxies of different morphologies might follow according to distinct physical processes taking place in each case. Quenching spiral galaxies are more likely undergoing secular processes, while elliptical galaxies have probably suffered more violent interactions such as major/minor mergers that might also affect their star formation histories. 

For this we use the Star Formation Acceleration (SFA) parameter that describes the recent variation in star formation rates, based on spectroscopic and photometric indices, and inferences from cosmological simulations. This allows us to directly measure the “quenching strength” of galaxies according to morphology, in addition to identifying those populations with bursting characteristics within the green valley, i.e. those with older stellar populations with a recent rejuvenation episode. 

Using SDSSDR12 and GALEX photometric and spectroscopic data and considering a morphological catalogue based on Deep Learning techniques (\citealp{barchi2019machine}), we analyzed the SFA distributions behavior, introduced by \cite{martin2017quenching}, for spirals and ellipticals in different color-bins. 

Our main results are as follows:

\begin{enumerate}
    \item SFA trends show that typical star formation histories within the green valley are statistically different for spirals and ellipticals. Moreover, we showed that ellipticals quench their star formation with shorter timescales than spirals. These results are in accordance with works in the local universe (e.g., \citealp{schawinski2014green}) and at intermediate redshifts ($z \sim 0.8$, \citealp{nogueira2018star}). This supports the scenario where minor/major mergers and violent interactions are responsible for rapidly quenching star formation and morphologically transforming galaxies (\citealp{di2007star}).
    
    \item SFA values within the blue cloud also tend to be different for spirals and ellipticals (with statistical relevance). This supports the scenario in which elliptical galaxies, even within the blue cloud, are not naturally star forming galaxies and already present quenching behavior that can be explained by a post merger object or a post rejuvenation process.
    
    \item A fraction of green valley galaxies are shown to be bursting, in accordance with \cite{martin2017quenching}. These objects are mostly massive and asymmetric spirals (Sections \ref{sub_results_sfa_mass} and \ref{sub_results_sfa_assy}, respectively). This result supports the scenario of major and minor mergers triggering star formation and changing the galaxy morphology. Lastly, we highlight that spirals can experience two evolutionary pathways, differentiated by the presence or absence of morphological disturbances.
\end{enumerate}

These results also present an exciting opportunity for future spectroscopic studies. Upcoming surveys such as the Dark Energy Spectroscopic Instrument (DESI; \citealp{aghamousa2016desi}) and future instruments such as MOONS at the VLT (\citealp{cirasuolo2016moons} - \citeyear{cirasuolo2020crescent}) and MOSAIC at the E-ELT (\citealp{hammer2016elt}) will allow us to obtain deeper spectra and absorption lines of $L_\star$ galaxies out to redshifts $z \sim 2-3$. This in turn would yield not only measurements of SFA in these objects, but the evolution of such trends as a function of morphology across cosmic time.



\vspace{2cm}

\textit{Acknowledgements:}
We would like to thank the anonymous referee for a productive report that helped clarify the results presented in this paper.

This study was financed in part by the Coordenação de Aperfeiçoamento de Pessoal de Nível Superior – Brasil (CAPES) – Finance Code 001. TSG would also like to thank the support of the National Research Council (Productivity in Research grant 314747/2020-6) and the Rio de Janeiro State Research Foundation (Young Scientist of Our State grant E-26/201.309/2021).

The material is based upon work supported by NASA under award number 80GSFC21M0002.

Funding for SDSS-III has been provided by the Alfred P. Sloan Foundation, the Participating Institutions, the National Science Foundation, and the U.S. Department of Energy Office of Science. The SDSS-III web site is http://www.sdss3.org/.

SDSS-III is managed by the Astrophysical Research Consortium for the Participating Institutions of the SDSS-III Collaboration including the University of Arizona, the Brazilian Participation Group, Brookhaven National Laboratory, Carnegie Mellon University, University of Florida, the French Participation Group, the German Participation Group, Harvard University, the Instituto de Astrofisica de Canarias, the Michigan State/Notre Dame/JINA Participation Group, Johns Hopkins University, Lawrence Berkeley National Laboratory, Max Planck Institute for Astrophysics, Max Planck Institute for Extraterrestrial Physics, New Mexico State University, New York University, Ohio State University, Pennsylvania State University, University of Portsmouth, Princeton University, the Spanish Participation Group, University of Tokyo, University of Utah, Vanderbilt University, University of Virginia, University of Washington, and Yale University. 

\section*{Data availability}

The image and spectroscopic data underlying this article are available in the SDSS-III database, at https://dx.doi.org/10.1088/0067-0049/219/1/12, and the GALEX database at https://dx.doi.org/10.1086/426387.
The SFA measurements and the Deep Learning morphological classification underlying this article will be shared on reasonable request to the corresponding author.




\bibliographystyle{mnras}
\bibliography{Bibliography} 

\appendix\section{\textit{NUV-r} vs sSFR}
\label{Appendix_sSFR}

Throughout the paper, we analyzed the SFA trends as a function of color, defining the blue cloud, green valley, and red sequence as proxies for star-forming, transition, and passive galaxies, respectively. In this appendix, we wish to investigate the results for SFA as a function of specific SFR (sSFR) and analyze how it compares to our main results. We use sSFR measurements from the MPA-JHU mass catalogs\footnote{https://www.sdss.org/dr14/spectro/galaxy\_mpajhu/}, based on the works by \cite{kauffmann2003stellar}, \cite{brinchmann2004physical}, and \cite{salim2007uv}. In order to define the blue cloud and red sequence for sSFR, we followed the method used for $NUV-r$, and delimited them considering $1\sigma$ of the blue and red peaks for its distribution. 

\begin{figure}
    \centering
    \includegraphics[width=\columnwidth]{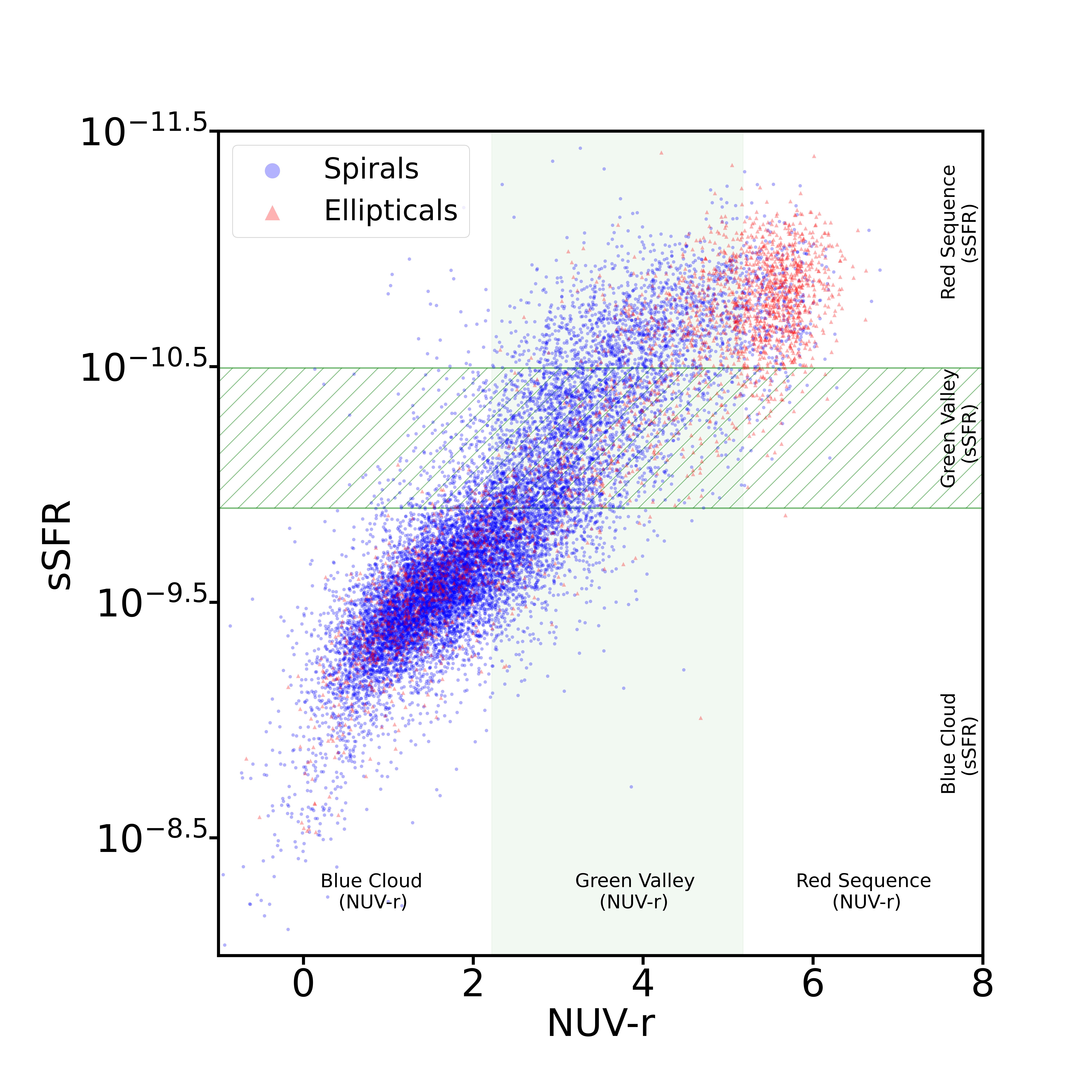}
    \caption{sSFR \textit{vs}. $NUV-r$ for spiral (blue) and elliptical (red) galaxies. We highlight the green valley region for each parameter, $NUV-r$ as a green vertical shaded area and sSFR as a green horizontal hatched area. The two quantities are strongly correlated, with similar populations in each region as defined by colour or sSFR. A fracion of $NUV-r$ green valley galaxies, however, are classified as red sequence according to their sSFR.} 
    \label{fig_nuvr_ssfr}
\end{figure}

Figure \ref{fig_nuvr_ssfr} shows a clear relation between these quantities, albeit with some dispersion. In Figure \ref{fig_nuvr_ssfr} we also show the regions for each parameter: blue cloud, green valley and red sequence. It is interesting to notice that the distributions for spirals and ellipticals into the regions are mostly consistent. However, a fraction of $NUV-r$ green valley galaxies are in the sSFR red sequence. These are probably galaxies that no longer form stars and are slowly evolving their color. With that we mind, we can follow for the analyses of SFA with sSFR. 

\begin{figure}
    \centering
    \includegraphics[width=\columnwidth]{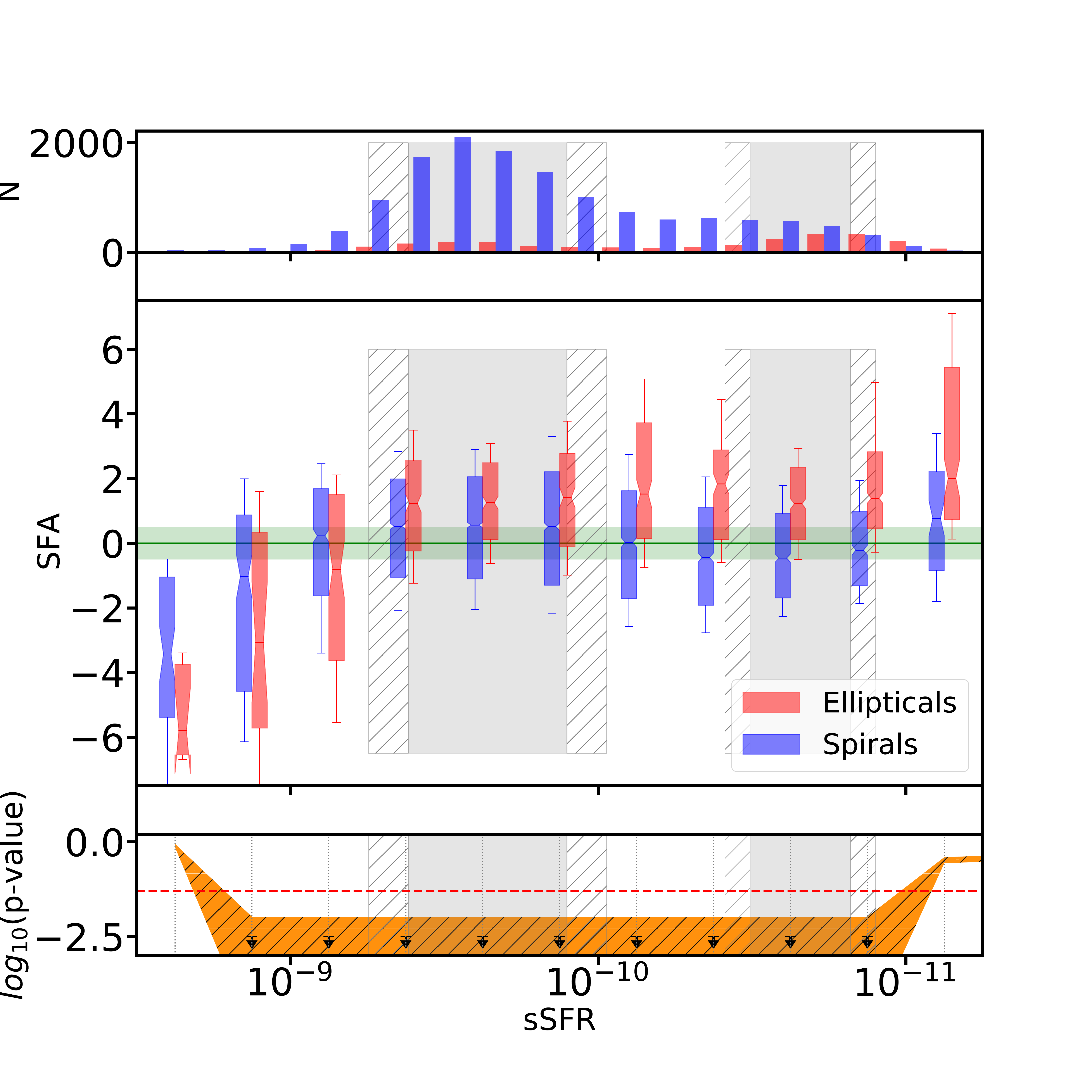}
    \caption{Same as Figure \ref{fig_sfa_morphology} but as a function of sSFR instead of $NUV-r$ for the x-axis.} 
    \label{fig_sfa_ssfr}
\end{figure}

In Figure \ref{fig_sfa_ssfr} we repeat the analysis shown in Figure \ref{fig_sfa_morphology} considering sSFR instead of $NUV-r$. We recover the same trend for the SFA dependence on morphology discussed in Sections \ref{chp4} and \ref{Chapter5}: SFA values for spirals smoothly evolve from negative to slightly positive as a function of sSFR, in accordance with a secular evolution scenario in which the galaxy quenches its star formation and slowly consumes the remainder of the gas reservoir. A significant fraction of green valley spirals also display display negative (\textit{bursting}) SFA values . The main difference lies in the red sequence, where values approach SFA$\sim 0$ as expected. This can be explained by Figure \ref{fig_nuvr_ssfr} and the dearth of red galaxies in our sample as discussed in previous sections. Additionally, ellipticals also show a similar trend: an abrupt transition of SFA values, from negative to positive, close to the blue cloud, and consistent quenching throughout the green valley. Lastly, we recover the statistical confirmation in the bottom panel of Figure \ref{fig_sfa_ssfr} that these populations are distinguishable. 

Overall, we conclude that the main results summarized on Section \ref{Summary} are robust and still stand when analysed as a function of sSFR instead of $NUV-r$ colours.

\bsp	
\label{lastpage}
\end{document}